\documentclass[fleqn,usenatbib,useAMS]{mnras}

\usepackage{graphicx}	
\usepackage{amsmath}	
\usepackage{amssymb}	
\usepackage{multicol}        
\usepackage{bm}		
\usepackage{pdflscape}	
\usepackage{xcolor}

\newcommand{\gx}{\textsc{Gadget-X}}

\newcommand{\ahf}{\textsc{AHF}}

\newcommand{\Mbnd}{{\ifmmode{M_{\rm bnd}}\else{$M_{\rm bnd}$}\fi}}
\newcommand{\Mfof}{{\ifmmode{M_{\rm fof}}\else{$M_{\rm fof}$}\fi}}
\newcommand{\Mcrit}{{\ifmmode{M_{\rm 200c}}\else{$M_{\rm 200c}$}\fi}}
\newcommand{\Rcrit}{{\ifmmode{R_{\rm 200c}}\else{$R_{\rm 200c}$}\fi}}
\newcommand{\Rhost}{{\ifmmode{R_{\rm host}}\else{$R_{\rm host}$}\fi}}
\newcommand{\Mmean}{{\ifmmode{M_{\rm 200m}}\else{$M_{\rm 200m}$}\fi}}
\newcommand{\MBN}{{\ifmmode{M_{\rm BN98}}\else{$M_{\rm BN98}$}\fi}}

\newcommand{\hGpc}{{\ifmmode{h^{-1}{\rm Gpc}}\else{$h^{-1}$Gpc}\fi}}
\newcommand{\hMpc}{{\ifmmode{h^{-1}{\rm Mpc}}\else{$h^{-1}$Mpc}\fi}}
\newcommand{\hkpc}{{\ifmmode{h^{-1}{\rm kpc}}\else{$h^{-1}$kpc}\fi}}
\newcommand{\hMsun}{{\ifmmode{h^{-1}{\rm {M_{\odot}}}}\else{$h^{-1}{\rm{M_{\odot}}}$}\fi}}
\newcommand{\Mstar}{{\ifmmode{M_{*}}\else{$M_{*}$}\fi}}
\newcommand{\Mhalo}{{\ifmmode{M_{\rm Halo}}\else{$M_{\rm Halo}$}\fi}}
\newcommand{\Ngal}{{\ifmmode{N_{\rm gal}}\else{$N_{\rm gal}$}\fi}}
\newcommand{\Norph}{{\ifmmode{N_{\rm orphan}}\else{$N_{\rm orphan}$}\fi}}
\newcommand{\Nxorph}{{\ifmmode{N_{\rm non-orphan}}\else{$N_{\rm non-orphan}$}\fi}}
\newcommand{\Zsolar}{{\ifmmode{Z_{\odot}}\else{$Z_{\odot}$}\fi}}
\newcommand{\Msun}{{\ifmmode{{\rm {M_{\odot}}}}\else{${\rm{M_{\odot}}}$}\fi}}
\newcommand{\ltsima}{$\; \buildrel < \over \sim \;$}
\newcommand{\gtsima}{$\; \buildrel > \over \sim \;$}
\newcommand{\lsim}{\lower.5ex\hbox{\ltsima}}
\newcommand{\gsim}{\lower.5ex\hbox{\gtsima}}

\newcommand{\Tab}[1]{Table~\ref{#1}}
\newcommand{\Sec}[1]{Section~\ref{#1}}

\newcommand{\Fig}[1]{Fig.~\ref{#1}}
\newcommand{\beq}{\begin{equation}}
\newcommand{\eeq}{\end{equation}}

\usepackage[T1]{fontenc}
\usepackage{ae,aecompl}

\usepackage{newtxtext,newtxmath}

\title[Galaxy pairs in simulated cluster environments]{Galaxy pairs in The Three Hundred simulations: a study on the performance of observational pair-finding techniques}

\author[Contreras-Santos et al.]
{Ana Contreras-Santos,$^{1}$\thanks{Contact e-mail: \href{mailto:ana.contreras@uam.es}{ana.contreras@uam.es}}
Alexander Knebe,$^{1,2,3}$
Weiguang Cui,$^{1,4}$
Roan Haggar,$^{5}$
\newauthor
Frazer Pearce,$^{5}$
Meghan Gray,$^{5}$
Marco De Petris,$^{6,7}$
and Gustavo Yepes$^{1,2}$
\\
$^{1}$Departamento de F\'isica Te\'{o}rica, M\'{o}dulo 15, Facultad de Ciencias, Universidad Aut\'{o}noma de Madrid, 28049 Madrid, Spain\\
$^{2}$Centro de Investigaci\'{o}n Avanzada en F\'isica Fundamental (CIAFF), Facultad de Ciencias, Universidad Aut\'{o}noma de Madrid, 28049 Madrid, Spain\\
$^{3}$International Centre for Radio Astronomy Research, University of Western Australia, 35 Stirling Highway, Crawley, Western Australia 6009, Australia\\
$^{4}$Institute for Astronomy, University of Edinburgh, Royal Observatory, Edinburgh EH9 3HJ, United Kingdom\\
$^{5}$School of Physics \& Astronomy, University of Nottingham, Nottingham NG7 2RD, United Kingdom\\
$^{6}$Dipartimento di Fisica, Sapienza Università di Roma, Piazzale Aldo Moro 5, 00185 Roma, Italy\\
$^{7}$I.N.A.F. - Osservatorio Astronomico di Roma, Via Frascati 33, 00040 Monteporzio Catone, Roma, Italy\\
}

\date{Last updated 2015 May 22; in original form 2013 September 5}

\pubyear{2021}

\begin{document}
\label{firstpage}
\pagerange{\pageref{firstpage}--\pageref{lastpage}}
\maketitle

\begin{abstract}
 Close pairs of galaxies have been broadly studied in the literature as a way to understand galaxy interactions and mergers. In observations they are usually defined by setting a maximum separation in the sky and in velocity along the line of sight, and finding galaxies within these ranges. However, when observing the sky, projection effects can affect the results, by creating spurious pairs that are not close in physical distance. In this work we mimic these observational techniques to find pairs in \textsc{The Three Hundred} simulations of clusters of galaxies. The galaxies' 3D coordinates are projected into 2D, with Hubble flow included for their line-of-sight velocities.
 The pairs found are classified into ``good'' or ``bad'' depending on whether their 3D separations are within the 2D spatial limit or not. We find that the fraction of good pairs can be between 30 and 60 per cent depending on the thresholds used in observations. 
 Studying the ratios of observable properties between the pair member galaxies, we find that the likelihood of a pair being ``good'' can be increased by around 40, 20 and 30 per cent if the given pair has, respectively, a mass ratio below 0.2, metallicity ratio above 0.8, or colour ratio below 0.8. Moreover, shape and stellar-to-halo mass ratios respectively below 0.4 and 0.2 can increase the likelihood by 50 to 100 per cent. These results suggest that these properties can be used to increase the chance of finding good pairs in observations of galaxy clusters and their environment.
\end{abstract}




\section{Introduction}

The $\Lambda$ cold dark matter ($\Lambda$CDM) growth paradigm for the Universe describes a hierarchical model of structure formation, where mergers between lower mass objects yield more massive systems \citep{White1978,Frenk2012}. In this context, interactions and mergers between galaxies are expected and these processes can play a very important role in galaxy formation and evolution. As such, several efforts have been devoted to studying the effects of  galaxy interactions that can result in property changes of the involved galaxies.

An essential quantity for studying the effects of mergers on galaxy formation and evolution is the fraction of galaxies that are undergoing a merger event. In observations, if high resolution images are available, galaxies undergoing mergers can be identified via their morphology \citep{Conselice2003,Lotz2004,Lotz2008,Lopez-Sanjuan2009}. Some studies even identify the stage of the merger the galaxy is in \citep{Pawlik2016}. However, this kind of data is not always acquirable, even more so if we want to study higher redshifts. A common approach taken at this point is to study galaxies that are close in the sky -- close pairs (e.g. \citealp{Carlberg1994,Patton2000,Kartaltepe2007,Duncan2019}). Although they do not directly trace mergers, close galaxies are more likely to merge in the near future. Assuming a certain merger timescale, the close-pair fraction can also be used to estimate the merger rate of galaxies \citep{Kitzbichler-White2008,Xu2012,Mundy2017}.

When compared to more isolated galaxies, galaxies with close companions have been shown to have enhanced star formation \citep{Barton2000,Li2008a,Scudder2012,Patton2013,Pan2018} and diluted metallicities \citep{Kewley2010,Rupke2010,Bustamante2020}. Some studies also show that close companions can induce the triggering of active galactic nuclei (AGN) \citep{Silverman2011,Cotini2013,Ellison2019}, although this has been a more debated topic, with some studies also failing to find a correlation between galaxy interactions and AGN enhancement \citep{Li2008b,Cisternas2011,Shah2020}. The dependence of these results on the environment the galaxies are in has also received some attention in the literature, suggesting that, in general, the tendencies are kept even when entering in high-density regions \citep{Perez2006a,Alonso2006,Alonso2012}.

To clarify this kind of matters and gain a better understanding of the topic, a more theoretical approach like the one given by numerical simulations can be very useful. 
Compared to observations, simulations have the advantage that they allow us to access different snapshots, and thus study what happens to the galaxies in a future or past moment. 
Previous works have studied the boost of star formation due to close companions (for example \citealt{Patton2020} in IllustrisTNG or \citealt{RodriguezMontero2019} in the Simba simulation). \citet{Bustamante2018} show that the metallicity dilution due to galaxy interactions is also seen in cosmological simulations. Idealized simulations of galaxy mergers have also found an increased AGN activity \citep{DiMatteo2005,Torrey2012}.

When trying to unify all these studies about close pairs, one problem that arises is regarding the selection criteria. In observations, galaxies are defined as ``close'' based on their separation in the sky and their separation in velocity along the line of sight. However, no unified threshold exists for these parameters to define a close pair. For the separation in the sky, the values range from 5 $\hkpc$ up to even 100 $\hkpc$, depending on the particulars of the study. For the velocity separation, the threshold is usually set to be between 300 and 1000 km/s difference. Studies like \citet{Husko2022} convert from one definition to another by assuming a power-law dependence of the close pair fraction on the maximum separation, but the comparisons remain difficult due to differing methodology.

In simulations, the 3D separation alone can be used to find close galaxies, although again the thresholds can differ depending on the study. By using directly the 3D coordinates, rather than 2D separation, simulations eliminate the problem of projection effects which can create spurious pairs of galaxies. Simulations also allow for checking if the two galaxies are in fact gravitationally bound, although we will not focus on that in this particular work.

The motivation for this project is to reduce this gap between observations and simulations, by answering the question of how likely it is that an observed galaxy pair (or group) in the sky is actually close in 3D. We aim to do this by using hydrodynamical simulations of galaxy clusters to mock the observational techniques used to find pairs in 2D. By measuring then the 3D separation between the galaxies in these pairs, we can assess the performance of these methods and thus the importance of the projection effects. While the effects of spurious pairs are expected to be stronger in high-density regions \citep{Mamon1986,Mamon1987}, a study by \citet{Alonso2004} found that real pairs dominate the statistics regardless of the environment, given the thresholds $r_\mathrm{sep} < 100$ $\hkpc$ in distance and $v_\mathrm{sep} < 350$ km/s in velocity. We intend to investigate this in more detail, exploring how it depends on the definition of ``close'' chosen, and if there are any properties of the pairs that influence it. We will also consider the circumstance of a galaxy having more than one close companion, i.e., being in a group, and similarly explore these situations.

In this work we are going to use \textsc{The Three Hundred} simulations\footnote{\url{https://the300-project.org/}}, which are 324 re-simulations of the most massive clusters in a dark matter only cosmological simulation. These clusters provide a high-density environment in where the interactions between galaxies are more frequent and can be especially important. On large scales, clusters are dominated by dark matter and hence gravity. But on smaller scales, also the interaction of the baryonic components of clusters plays an important role (see \citealp{Kravtsov2012} for a review on galaxy clusters). This leads to several different phenomena that drive galaxy evolution, also making clusters very interesting environments to study galaxy interactions \citep{Gnedin2003,Park2009,Boselli2014}. 
Using \textsc{The Three Hundred} simulations allows us to have lots of statistics, with many pairs and groups to study, belonging to clusters with different properties, and hence different environments. 

The paper is organized as follows. In \Sec{sec:data}, we present the details of the simulation and the halo catalogues used to identify the haloes. In \Sec{sec:method} we present the method used to find close pairs of galaxies. For the pairs found this way, in \Sec{sec:stats2D} we present some statistics regarding the number of pairs found. In \Sec{sec:results1} we separate our projected pairs into ``good'' and ``bad'' by measuring their 3D distance compared to the 2D one. In the following section, \Sec{sec:results2}, we study if there is any correlation between observational properties of the pairs and their ``goodness''. Finally, in \Sec{sec:conclusions}, we summarize and discuss our results.


\section{The Data} \label{sec:data}

\subsection{The Three Hundred Clusters}
The clusters in \textsc{The Three Hundred} data set were created upon the DM-only MDPL2 MultiDark Simulation\footnote{The MultiDark simulations – incl. the MDPL2 used here – are publicly available at \url{https://www.cosmosim.org}} \citep{Klypin16}. This simulation is a periodic cube of comoving length 1 $\hGpc$ containing $3840^3$ DM particles, each of mass $1.5\times 10^9$ $\hMsun$, with cosmological parameters based on the Planck 2015 cosmology \citep{Planck2015}. From this simulation, the 324 clusters with the largest halo virial mass\footnote{The halo virial mass is defined as the mass enclosed inside an overdensity of $\sim$98 times the critical density of the universe \citep{Bryan98}} at $z=0$ were selected. These clusters serve as the centre of spherical regions with radius 15 $\hMpc$, where the initial DM particles were traced back to their initial conditions and then split into dark matter and gas particles according to the cosmological baryon fraction; leading to a dark matter and gas particle mass resolution of $m_\mathrm{DM}=1.27 \times 10^9$ $\hMsun$ and $m_\mathrm{gas}=2.36 \times 10^8$ $\hMsun$, respectively. Outside these regions, dark matter particles were degraded with lower mass resolution particles to reduce the computational cost but keeping large scale tidal effects. From these initial conditions, each cluster region was then re-simulated including now full hydrodynamics with the SPH code \gx, using a Plummer equivalent softening of 6.5 $\hkpc$ for both the dark matter and baryonic component. The output includes, for each of the 324 clusters, 129 snapshots between $z=16.98$ and $z=0$. At $z=0$, the 324 central galaxy clusters have a mass range from $M_{200} = 6.4 \times 10^{14} \hMsun$ to $M_{200} = 2.65 \times 10^{15} \hMsun$. \textsc{The Three Hundred} data set was presented in an introductory paper by \citet{Cui18}, and several other papers have been published based on this data (see e.g. \citealp{Wang18,Mostoghiu18,Haggar2020}), to which we refer the reader for more details about this project.

\gx, the code used for the re-simulations, is a modified version of the non-public \textsc{Gadget3} code \citep{Murante2010,Rasia2015,Planelles2017,Biffi2017}, which evolves dark matter with the \textsc{Gadget3} Tree-PM gravity solver (an advanced version of the \textsc{Gadget2} code; \citealp{springel_gadget2_2005}). It uses an improved SPH scheme that includes artificial thermal diffusion, time-dependent artificial viscosity, high-order Wendland C4 interpolating kernel and wake-up scheme (see \citealp{Beck2016} and \citealp{Sembolini2016} for a presentation of the performance of this SPH algorithm). Star formation is carried out as in \citet{Tornatore2007}, and follows the star formation algorithm presented in \citet{Springel03}. Black hole (BH) growth and AGN feedback are implemented following \citet{Steinborn2015}, where super massive black holes (SMBHs) grow via Bondi-Hoyle like gas accretion (Eddington limited), with the model distinguishing between a cold and a hot component.

\subsection{The Halo Catalogues}
The halo analysis was done using the open-source AHF halo finder \citep{Gill04a,Knollmann09}, which includes both gas and stars in the halo finding process. Haloes, as well as substructures, are found by locating overdensities in an adaptively smoothed density field (see e.g. \citealp{Knebe11} for more details on halo finders). AHF computes the radius $R_{200}$ of each halo identified, which is the radius $r$ at which the density $\rho(r)=M(<r)/(4\pi r^3/3)$ drops below 200 times the critical density of the Universe at the given redshift, $\rho_{\mathrm{crit}}$. $R_{500}$ is defined accordingly, together with the corresponding enclosed masses $M_{200}$ and $M_{500}$. Subhaloes are defined as haloes which lie within the $R_{200}$ region of a more massive halo, the so-called host halo. This way, the mass of this host halo includes the masses of all the subhaloes contained within it.

AHF also allows other properties to be generated for each halo and subhalo
. The luminosity (and magnitude) in any spectral band is calculated by applying the stellar population synthesis code \textsc{stardust} (see \citealp{Devriendt99}, and references therein for more details). This code uses the given age and metallicity of each stellar particle within a given halo and obtains from a catalogue the full spectrum of a star with those properties. Then, each spectrum is weighted by the mass of the particle, and the sum of all of them yields the galaxy SED. A Kennicutt initial mass function \citep{Kennicutt98} is assumed for these calculations.

To trace the haloes through the different snapshots of the simulations, merger trees are built using the tree-builder \textsc{MergerTree}, which comes with the AHF package. However, in this work we will only work with the $z=0$ snapshots.

\section{Methodology} \label{sec:method}
In this Section we first present the way observers find close pairs of galaxies in the sky. Then we explain how we apply this same method to \textsc{The Three Hundred} data set, and how we intend to assess its limitations by using the simulations' data.

\subsection{Finding pairs in observations}
In observational studies, the way of finding close pairs is  based on two quantities that relate two different galaxies: their projected separation in the sky and the separation in velocity along the line of sight. Defining a maximum threshold for these two quantities, $r_\mathrm{sep}$ and $v_\mathrm{sep}$, it can be determined if two galaxies are close. After applying some kind of selection criterion for the galaxies, e.g. based on their luminosity, the pairs can be easily found by checking if these parameters are below the maximum set. In the literature, the values used as thresholds depend a lot on the specific purpose of the study, so that works focusing more on mergers themselves use smaller separations (e.g. \citet{Robotham2014} use $r_\mathrm{sep} = 20$ kpc/h and $v_\mathrm{sep} = 500$ km/s), while works on the effects of interactions and companions can use separations up to the order of Mpc (e.g. in \citet{Patton2016} they use $r_\mathrm{sep} = 2$ Mpc and $v_\mathrm{sep} = 1000$ km/s).

\subsection{Application to simulations}
We will now apply the same observational method to our simulations, trying to mimic the procedure as much as possible.
Before doing any further calculation, we will first make a selection of the objects we are going to work with from the simulations. In this work, we will use the word ``galaxy'' to refer to the objects in the hydrodynamical simulations, including both their stellar and dark matter components. For each of the 324 clusters in \textsc{The Three Hundred} data set, we first select the region within $5R_{200}$ of the cluster centre, to also include the cluster outskirts. To all the galaxies within this region, we apply a stellar mass cut similarly to observational studies, so that we work only with the galaxies with $M_\mathrm{*}>10^{9.5} \hMsun$ (see \citealp{Cui18} for the stellar mass function of all the galaxies in our data set). 
This way we end up with a total number of galaxies between $\sim 400$ and 1200 depending on the cluster, which is still statistically high enough.
Note that, throughout this work we only use the AHF catalogues at $z=0$, i.e. only the last snapshot of the simulations.

Once we have done the galaxy selection, we can proceed to find the pairs between them. In order to produce results that are comparable to observations, we create different 2D projections of the 3D coordinates. We do this by rotating the coordinates about two orthogonal random axes after placing the cluster at the centre of the coordinate system, and then projecting the coordinates into a 2D plane. This way, we obtain 100 random projections for each cluster, to which we can apply our pair-finding algorithm, based on the two parameters $r_\mathrm{sep}$ and $v_\mathrm{sep}$. The separation between two galaxies is simply computed as their distance in the 2D plane, while the velocity separation comes from two contributions: the peculiar velocity of the galaxies along the remaining axis and the difference in recession velocities due to the Hubble flow. To include both terms in our calculations, before doing the projections we add to the peculiar velocity (given by AHF) the contribution of the Hubble flow as $H \cdot \textbf{r}$, where $H$ is the Hubble constant and $\textbf{r}$ are the coordinates of each object. 
Then, after the coordinates are projected, the velocity separation is simply the difference in velocities along the line-of-sight axis.

In order to study the dependence of the results on the values chosen, and to have some flexibility in our pair definition, we have decided to use several different values for the thresholds $r_\mathrm{sep}$ and $v_\mathrm{sep}$ that determine if two galaxies are close. Based on the literature (see e.g. \citealp{Husko2022} for a summary of different studies) we have used three different values for each of the parameters, which can thus be combined in 9 different ways. For the separation in the sky we use the values: 20, 50 and 100 kpc/h, while for the velocity separation we use 300, 500 and 1000 km/s. 

Finally, to be able to assess if a galaxy has more than one close companion, we also allow for groups to be created connecting the galaxies. This way, if there are more than two galaxies grouped together, we will define that as a group rather than a pair. Regarding this, our algorithm links galaxies (or pairs) with existing pairs by establishing friends-of-friends connections. This means that if there is a pair of close galaxies A and B, and galaxy C is found to be close to only one of them, the three of them will still be classified as a group. Then, if a galaxy D is close to any of the three galaxies, it will also be included in the group, and so on.

\begin{figure}
  \hspace*{-0.1cm}
  {\includegraphics[width=8.5cm]{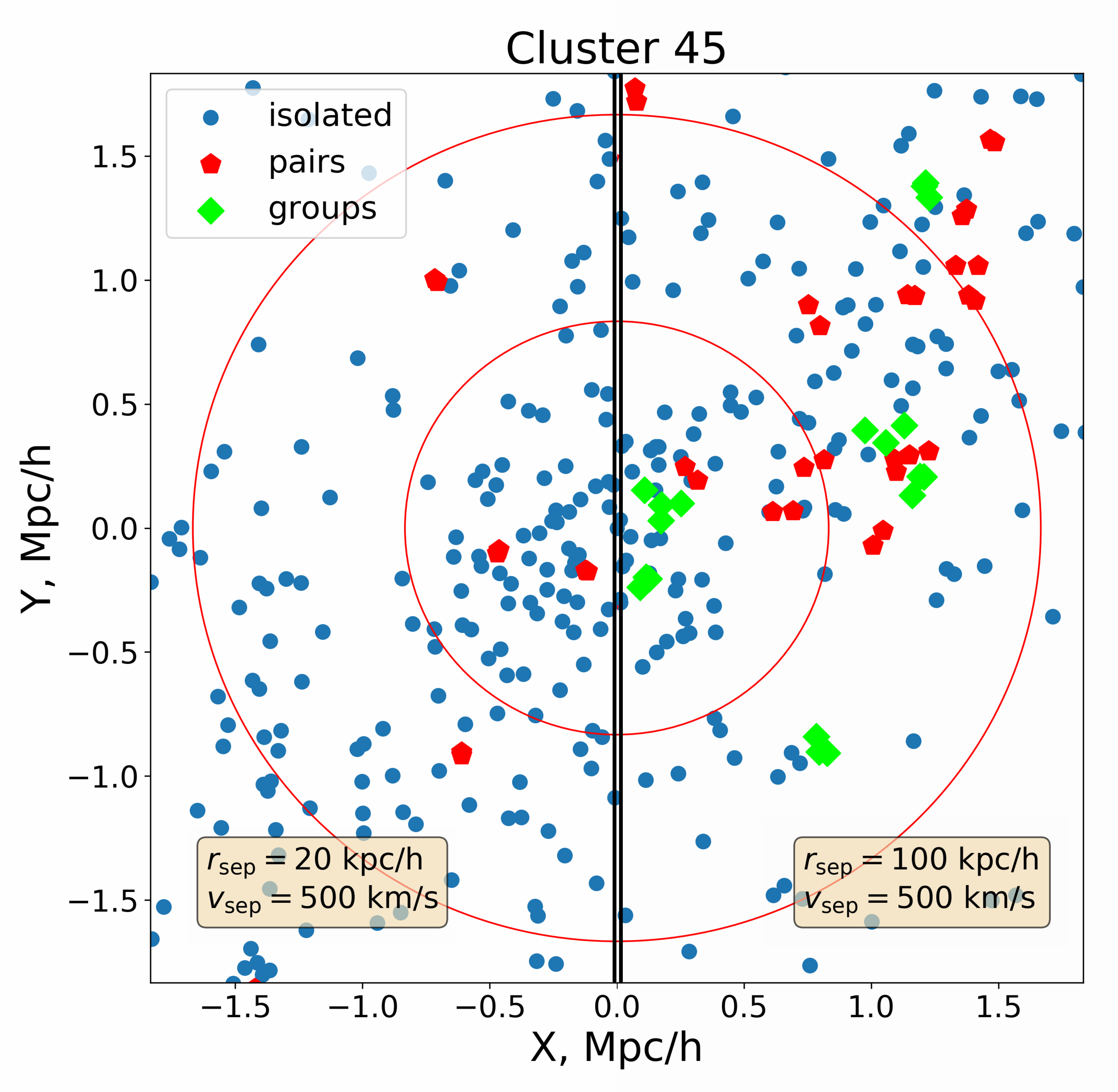}}
  \caption{Example of one random projection of the $R_{200}$ region of cluster 45 in \textsc{The Three Hundred} data set. The blue dots are isolated galaxies, the red pentagons are galaxies in pairs, and the green diamonds represent those in groups (with three or more members). For comparison, the plot is divided in half, with the definition of close used for each half indicated in the corresponding corner. The two circles delimitate the $R_{200}$ and $0.5 \cdot R_{200}$ regions. The axes are centred so that the cluster centre is at $X=0$, $Y=0$ Mpc/h.}
\label{fig:method_reg45}
\end{figure}

\Fig{fig:method_reg45} shows an example of the pairs and groups found for one random 2D projection for cluster 45 in \textsc{The Three Hundred} sample, for two different definitions of close. In the left for $r_\mathrm{sep} = 20$ kpc/h and in the right for $r_\mathrm{sep} = 100$ kpc/h, while $v_\mathrm{sep}$ was set at $500$ km/s for both of them. The blue dots show the isolated galaxies, the red pentagons are the galaxies in pairs and the green diamonds are those in groups (with three or more members). Comparing both halves in \Fig{fig:method_reg45} we can see the effects of increasing the maximum separation, finding not only more pairs but also a significant amount of groups. 

Once the pairs and groups are found for all the clusters and projections, they can be analysed in terms of their statistics and properties, making use of all the information available in the simulations. 
For that, we will stack together these results (distribution and properties of pairs and groups) for the 100 projections for each of the 324 clusters. This way, in our results we can distinguish between the clusters but not between projections for the same cluster.


\subsubsection{Defining 2D pairs as `good'} \label{sec:method-class}
Using the previous methodology we can order all the projected galaxies into different pairs and groups. Then, taking advantage of the power of simulations, the 3D coordinates of the galaxies can be used to determine if the projected pairs are also close in real space. For simplicity, for this part of the study we will focus only on galaxy pairs. We will see in the following section (\Sec{sec:stats2D}) that we are not losing much information by doing this, since pairs dominate the statistics.

In order to address the question of the ``goodness'' of a pair, we use the physical (3D) separation between the involved galaxies. Our idea is to check if the 3D separation between the galaxies is within the range allowed by the $r_\mathrm{sep}$ threshold in projected distance. To quantify this in a single parameter we compute the ratio $r_\mathrm{3D}/r_\mathrm{sep}$, where $r_\mathrm{3D}$ is the physical separation between the galaxies in the simulation and $r_\mathrm{sep}$ is fixed to 20, 50 or 100 kpc/h, depending on the value chosen for the pair-finding method. Using this ratio, we can simply set a threshold at $r_\mathrm{3D}/r_\mathrm{sep} = 1$, so that pairs with this ratio below (or equal to) 1 can be tagged as ``good'' pairs, with a 3D separation within the allowed range; and pairs with $r_\mathrm{3D}/r_\mathrm{sep} > 1$ are classified as ``bad'' pairs. Note that we call them ``good'' and ``bad'' pairs, since our criterion is just based on the projected and physical separations between the galaxies in the pair (rather than e.g. ``true'' and ``false''), and thus it only allows us to assess the degree to which the 3D distance is within the maximum separation allowed. Other criteria, like the one used in \citet{Haggar2021}, can be used to determine if the two galaxies are gravitationally bound, but for this work we prefer to focus on this simple comparison, which already gives a significant amount of knowledge about the pairs. We will take a different approach and investigate a separation between ``true'' and ``false'' pairs in a follow-up work (Contreras-Santos et al., in prep.). 

In the following sections we will first show some statistics of the way the galaxies are connected using the methodology described, regarding the number of pairs and groups found and the galaxies that belong to them. Then, for the pairs, we will separate them into ``good'' and ``bad'' according to the criterion we just described. We will study the fraction of pairs that belong to each class and analyse their properties based on this division. 

\section{Statistics of 2D grouping} \label{sec:stats2D}
Using the methodology described in the previous section, we can link the galaxies to their close projected companions in each of the 324 clusters. In this section we will analyse the statistics of the pairs and groups found, by computing the fraction of galaxies that are in a pair or group, the number of members that each association of galaxies has (i.e. if they are pairs or groups) and finally by seeing if there is any dependence of these results on the properties of the specific cluster.

\begin{figure*}
  \hspace*{-0.1cm}
  \includegraphics[width=16cm]{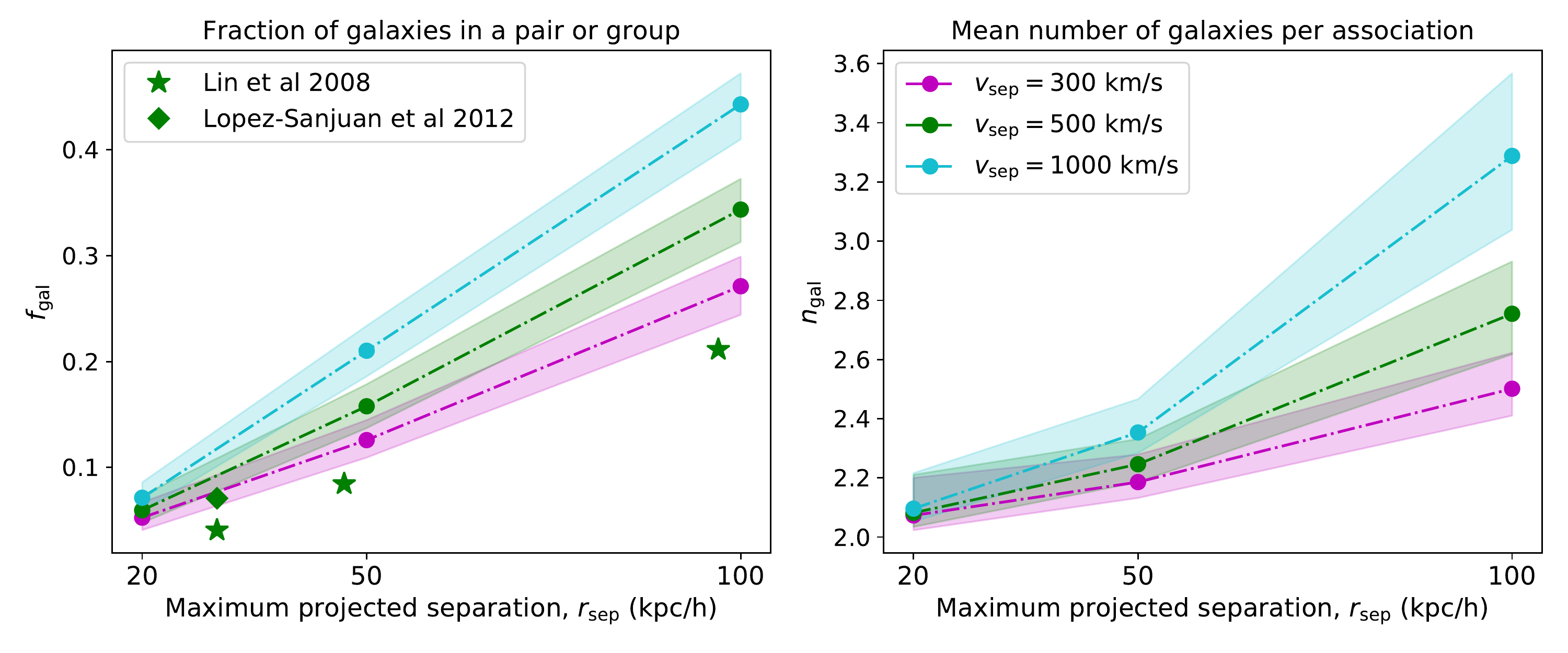}
  \vspace*{-0.2cm}
  \caption{\textbf{Left}, fraction of galaxies that are in a group or pair, i.e. that have at least one close companion, as a function of maximum separation chosen, $r_\mathrm{sep}$. The dots show the median values of the distribution for the 324 clusters in \textsc{The Three Hundred} data set, with the shaded regions indicating the 16-84 percentiles. For each cluster, the value is computed as the mean percentage for 100 random projections. The colours indicate the velocity threshold $v_\mathrm{sep}$ used, in magenta 300 km/s, in green 500 km/s and in cyan 1000 km/s. Previous studies have also been included, as indicated in the legend.
  \textbf{Right}, same as the left panel but for the mean number of members per galaxy association (see explanation in the text).}
\label{fig:stats2d}
\end{figure*}

\subsection{Fraction of galaxies in pairs and groups}
For each of the 324 clusters in \textsc{The Three Hundred} data set, we connect the galaxies as described in \Sec{sec:method} and count the fraction of them that remain in a pair or group (in 2D), i.e., the fraction of galaxies that have at least one close companion. We do this for the 100 random 2D projections for each cluster. This way we can obtain the mean fraction for each cluster, together with the standard deviation. By putting all the mean values together, we can compute the median value of this distribution, together with its 16-84 percentiles. This is shown in the left panel of \Fig{fig:stats2d} (dots for the median values and shaded regions for the percentiles). To compare the results, we show this median value as a function of the maximum projected separation used, $r_\mathrm{sep}$, and for the three different values of the velocity separation used, $v_\mathrm{sep}$, as indicated in the legend.

In the plot we can see that the results show a strong dependence on the chosen value of the separation. For the lowest threshold $r_\mathrm{sep} = 20$ kpc/h, the fraction is below 8 per cent, with very small dependence on the velocity threshold. For the higher values of $r_\mathrm{sep}$ the velocity threshold starts to make a difference. The fraction of galaxies with a close companion is between $\sim 12$ and 21 per cent for 50 kpc/h separation, while for 100 kpc/h the range becomes even greater, going from $\sim 27$ for $v_\mathrm{sep} = 300$ km/s (in magenta) to more than 40 per cent for $v_\mathrm{sep} = 1000$ km/s (in cyan).

We also include some previous observational results in the left panel of \Fig{fig:stats2d}. The green diamond is extracted from \citet{Lopez-Sanjuan2012}, where data from the COSMOS survey are analysed. They use the thresholds $r_\mathrm{sep}=30$ kpc/h and $v_\mathrm{sep}=500$ km/s to measure the close pair fraction for galaxies with $M_\mathrm{*}\geq 10^{11} \Msun$. In spite of this different selection criterion, we can see that, although a bit smaller, their value is compatible with the expected result for our data. We also show, as stars, the results from \citet{Lin2008}, where they study the pair fraction of galaxies from the DEEP2 Redshift Survey, applying a cut by luminosity. In this case, they employ three different maximum separations, $r_\mathrm{sep}=30$, 50 and 100 kpc/h, with $v_\mathrm{sep}=500$ km/s. We see that their results are in general smaller than ours, even compared to our $v_\mathrm{sep}=300$ km/s results. However, we have to note here that \citet{Lin2008} results come from a wide survey rather than a cluster environment like ours, where the galaxy density is higher. In this same work by \citet{Lin2008}, and also in later works \citep{Lin2010,deRavel2011}, the pair fraction is found to be dependent on the galaxy number density, so that our results are in fact expected to be higher, as seen in \Fig{fig:stats2d}.

\subsection{Number of members per galaxy association} 
In addition to the percentage of galaxies that have at least one close companion, i.e. are in a pair or group, it can also be of interest to know the size of these associations of galaxies. Are the galaxies mainly in pairs or in bigger groups? Similarly to the previous subsection, we have computed the mean number of galaxies per association (we use the word ``association'' here to refer to pairs and groups at the same time), again for each of the 324 clusters as the mean value for the 100 projections. In the right panel of \Fig{fig:stats2d} we show the median and 16-84 percentiles of the distribution of this value for the 324 clusters. The format is the same as in the left panel, with the X-axis indicating the maximum separation used and the colours indicating the velocity threshold. Looking first at the minimum value of $r_\mathrm{sep}$, we see that the mean number of galaxies is very slightly above 2 (regardless of the velocity), meaning that almost all non-isolated galaxies are in pairs, with very few groups. In fact, we find that groups represent around 7 per cent of all the galaxy associations for this $r_\mathrm{sep}$ value. For $r_\mathrm{sep}=50$ kpc/h the increase is less significant than in the left panel of \Fig{fig:stats2d}, with the number ranging between $2.2$ and $2.4$. This means that for this separation the galaxy associations are still very predominantly pairs, the velocity threshold still not being too determining. The percentage of groups is now between 15 and 20 per cent of all the associations found. Finally, for $r_\mathrm{sep}=100$ kpc/h, the median of the mean number of galaxies is $2.5$ for $v_\mathrm{sep}=300$ km/s (in magenta), showing that galaxy pairs are still predominant but more groups are found now. We can see that the value increases notably with the maximum velocity allowed, the mean number of galaxies even reaching $3.2$ for $v_\mathrm{sep}=1000$ km/s, although the scatter in this value is also becoming bigger. In this case, the percentage of groups ranges between $\sim$ 25 and 35 per cent for the different velocity thresholds, showing that pairs dominate the statistics even for the least restrictive thresholds.

In this case we limit the plot to our own results, since this quantity has not been that thoroughly studied in other works and thus the comparison is more difficult. However we believe the results in the right panel of \Fig{fig:stats2d} are important by themselves, to see how the values of $r_\mathrm{sep}$ and $v_\mathrm{sep}$ affect the way the galaxies are ordered in groups and pairs, and to gain a better understanding of the results shown throughout this work.


\subsection{Dependence of the 2D statistics on cluster properties} \label{sec:stats_corr}
In this subsection we study the correlation of the previous statistics, the fraction of galaxies in a pair or group, $f_\mathrm{gal}$ and the mean number of members per galaxy association, $n_\mathrm{gal}$, with two different cluster properties: its mass and dynamical state, which can be derived in observational studies, too.
This way, we can investigate if these two properties of clusters have an effect on the distribution of projected galaxies in pairs and groups, an information that can be useful for future studies and to gain a better understanding of close pairs of galaxies in cluster environments. 

\subsubsection{Mass}
In the top panel in \Fig{fig:stats_corr} we show the correlation between the fraction of galaxies that are found to be in a pair or group and the mass of the cluster. We use the mass $M_{200}$ at $z=0$, given directly by AHF. For simplicity, we only show the results using $r_\mathrm{sep}=100$ kpc/h and $v_\mathrm{sep}=500$ km/s to link the galaxies, but we will comment on the results for different values. In the plot we can see that there is no correlation at all, which could be expected since to obtain this fraction we are already normalising by the total number of galaxies. For different definitions of proximity, i.e. different values of $r_\mathrm{sep}$ and $v_\mathrm{sep}$ the results are the same, with essentially no dependence of the results on the mass. Although in the plot we only show the results for the fraction of galaxies with at least one close companion, the conclusion still holds for the mean number of galaxies per association, with no correlation with the mass found.

\subsubsection{Dynamical state}
To quantify the dynamical state of clusters we use the so-called `relaxation parameter', introduced by \citet{Haggar2020} as:
\begin{equation}
\chi_{\mathrm{DS}}=\sqrt{\frac{3}{\left(\frac{\Delta_r}{0.04}\right)^2+\left(\frac{f_s}{0.1}\right)^2+\left(\frac{\vert 1 - \eta \vert}{0.15}\right)^2}}.
\label{eq:chiDS}
\end{equation}
This equation is based on the three parameters initially introduced by \citet{Neto07} as proxies for relaxation. Later studies by \citet{Cui18} or \citet{DeLuca2021} also used these parameters to study the relaxation of the clusters in \textsc{The Three Hundred} sample. The parameters are the \textbf{centre of mass offset}, $\Delta_r$, which is the offset between the positions of the centre of mass of the cluster and the density peak, normalised to the halo radius; the \textbf{subhalo mass fraction}, $f_s$, which is the fraction of cluster mass contained in subhaloes; and the \textbf{virial ratio} $\eta$ \citep[see][for an updated calculation for hydrodynamic simulations]{Cui2017}. For a cluster to be most relaxed, $\Delta_r$ and $f_s$ have to be minimised, and $\eta \rightarrow 1$, and so `dynamically relaxed' clusters have $\chi_{\mathrm{DS}} \gtrsim 1$. The relaxation parameter $\chi_\mathrm{DS}$ thus provides a continuous way to classify the dynamical state of the clusters.

\begin{figure}
   \hspace*{-0.1cm}\includegraphics[width=8.5cm]{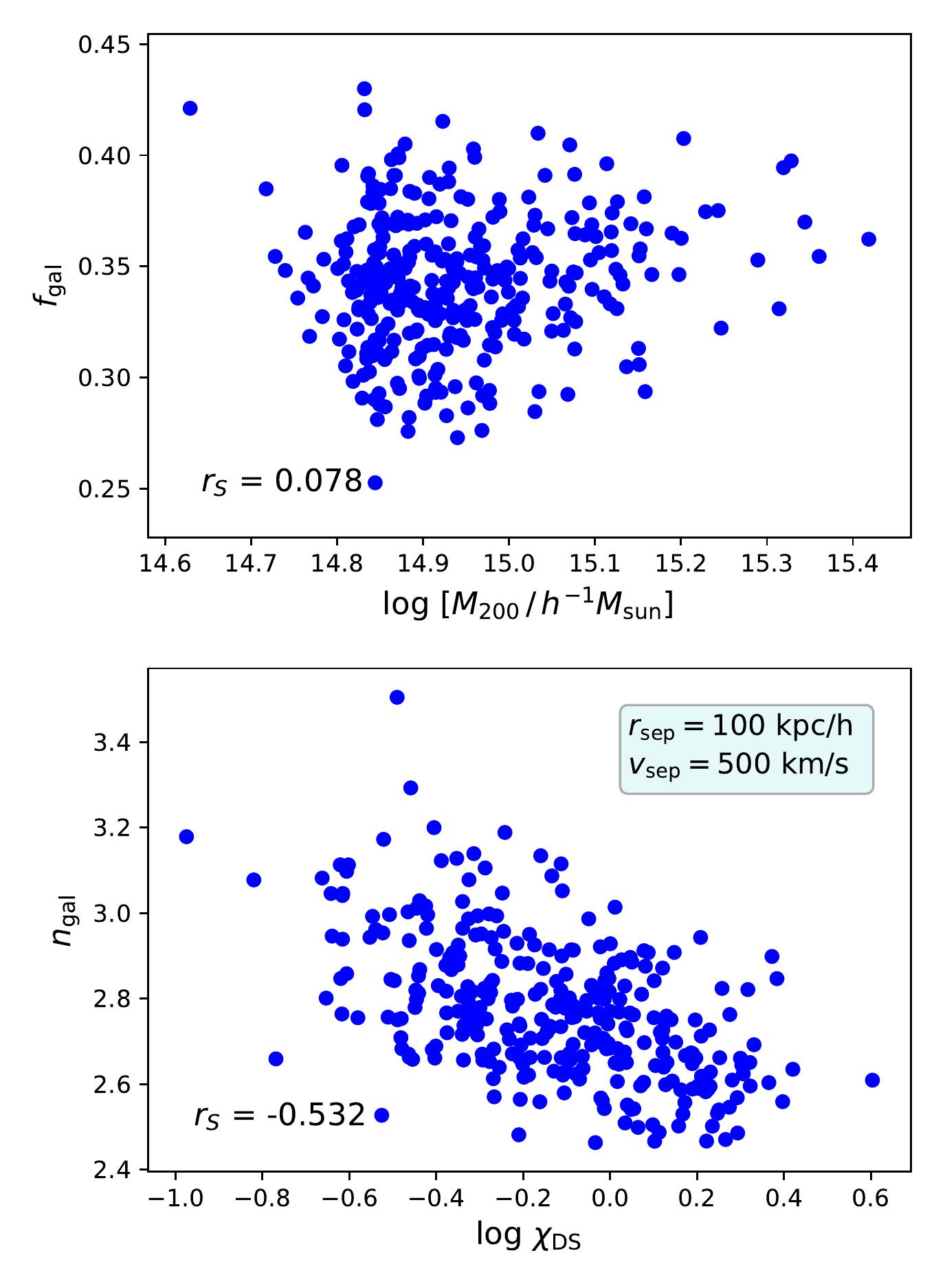}
   \caption{\textbf{Top}, fraction of galaxies in a group, computed for $r_\mathrm{sep}=100$ kpc/h and $v_\mathrm{sep}=500$ km/s, as a function of mass of the cluster, in logarithmic scale.
  \textbf{Bottom}, mean number of galaxies per group as a function of the relaxation parameter $\chi_\mathrm{DS}$, also in logarithmic scale. In both panels, the Spearman correlation coefficient, $r_S$, is indicated in the bottom left corner.}
\label{fig:stats_corr}
\end{figure}

In the lower panel of \Fig{fig:stats_corr} we show a scatter plot of the mean number of galaxies per association against the relaxation parameter at $z=0$. As with the mass, we only show the results for $r_\mathrm{sep}=100$ kpc/h and $v_\mathrm{sep}=500$ km/s. In this case we see that there is a clear negative correlation, meaning that in more disturbed clusters the associations of galaxies are more likely to be bigger. Although not shown here, we see a very similar correlation with the fraction of galaxies in a pair or group, $f_\mathrm{gal}$, meaning that galaxies are less likely to be isolated in these disturbed clusters. These results are in line with what could be expected, since relaxed clusters have less substructure it is logical for them to have fewer pairs and groups. The value of the Spearman correlation coefficient is $r_S \lesssim -0.5$ in both cases, and remains the same when changing $v_\mathrm{sep}$ for the same separation threshold. However, the strength of the correlation decreases if we use lower values of $r_\mathrm{sep}$, even disappearing for the most `restrictive' definition with $r_\mathrm{sep}=20$ kpc/h and $v_\mathrm{sep}=300$ km/s. So in general we can say there is a negative correlation with the dynamical state of the cluster, so that more relaxed clusters have less galaxies connected to other galaxies and the groups are smaller, but this does not hold for all the possible definitions of galaxy pair.

\section{Fraction of `good' 2D pairs} \label{sec:results1}

Now that we have studied in a quantitative way the projected pairs and groups of close galaxies found following the methodology in \Sec{sec:method}, we can focus the study on the pairs themselves, concentrating on their particular properties. As a reminder, the objective of the work presented here is the quantification of the likelihood for a pair seen in 2D to be actually close in 3D. As it has also been said before, we will now work only with pairs, i.e. only with the two-member groups, since in \Sec{sec:stats2D} we showed that they are very predominant in most of the cases, and they are also much easier to analyse. 

In order to assess the ``goodness'' of a pair, we use the method described in \Sec{sec:method-class}, based on $r_\mathrm{3D}/r_\mathrm{sep}$, the ratio of the 3D distance to the maximum 2D separation allowed. We set the threshold for classifying the pairs at 1, so that ``good'' pairs are those with $r_\mathrm{3D} \leq r_\mathrm{sep}$, with $r_\mathrm{sep}$ being fixed for the given definition. Here we will show some of the results obtained for the given pairs.

\begin{figure}
   \hspace*{-0.1cm}\includegraphics[width=8.5cm]{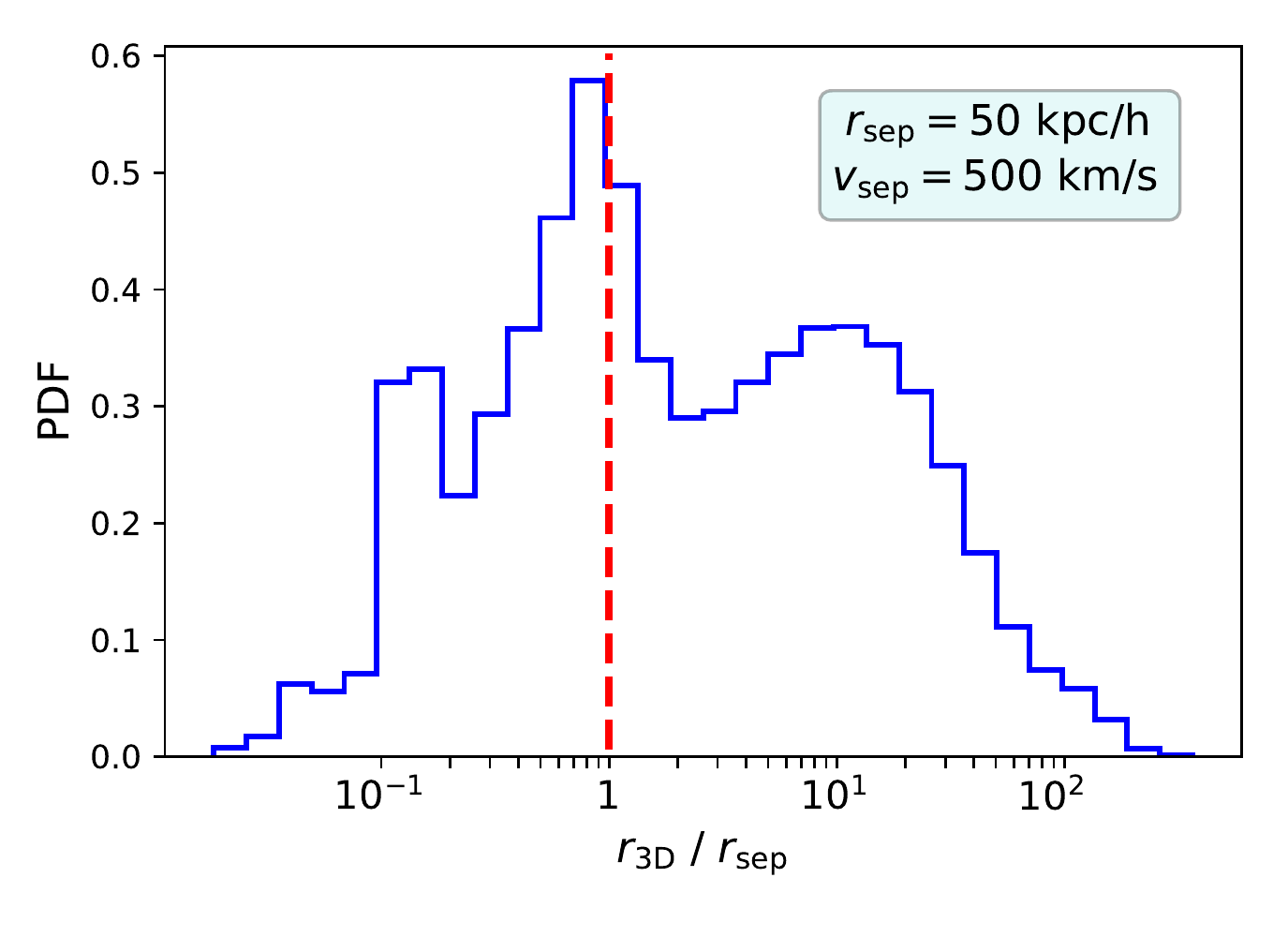}
   \caption{Distribution of the $r_\mathrm{3D}/r_\mathrm{sep}$ ratio for all the pairs found with $r_\mathrm{sep}=50$ kpc/h and $v_\mathrm{sep}=500$ km/s, stacking all the 324 clusters together. The red dashed line indicates the threshold used to separate good and pairs, $r_\mathrm{3D}/r_\mathrm{sep}=1$.} 
\label{fig:rsep3dratio}
\end{figure}

In \Fig{fig:rsep3dratio} we show the distribution of the ratio $r_\mathrm{3D}/r_\mathrm{sep}$ for all the pairs found for the given definition $r_\mathrm{sep}=50$ kpc/h and $v_\mathrm{sep}=500$ km/s. We stacked all the pairs together, regardless of the cluster. We can see that the distribution shows a peak at $\sim 1$, but extends to quite higher values of the ratio, corresponding to pairs that are much farther away in 3D distance than allowed by the $r_\mathrm{sep}$ criterion. For simplicity, in this plot we only show the results for one definition, but the conclusions hold for all of them. In the plot we have indicated this threshold at $r_\mathrm{3D}=r_\mathrm{sep}$ with a red dashed line, that separates the distribution into ``good'' (to the left of this value) and ``bad'' (to the right) pairs. This way, we can easily measure the fraction of good pairs.

\begin{figure}
   \hspace*{-0.1cm}\includegraphics[width=8.5cm]{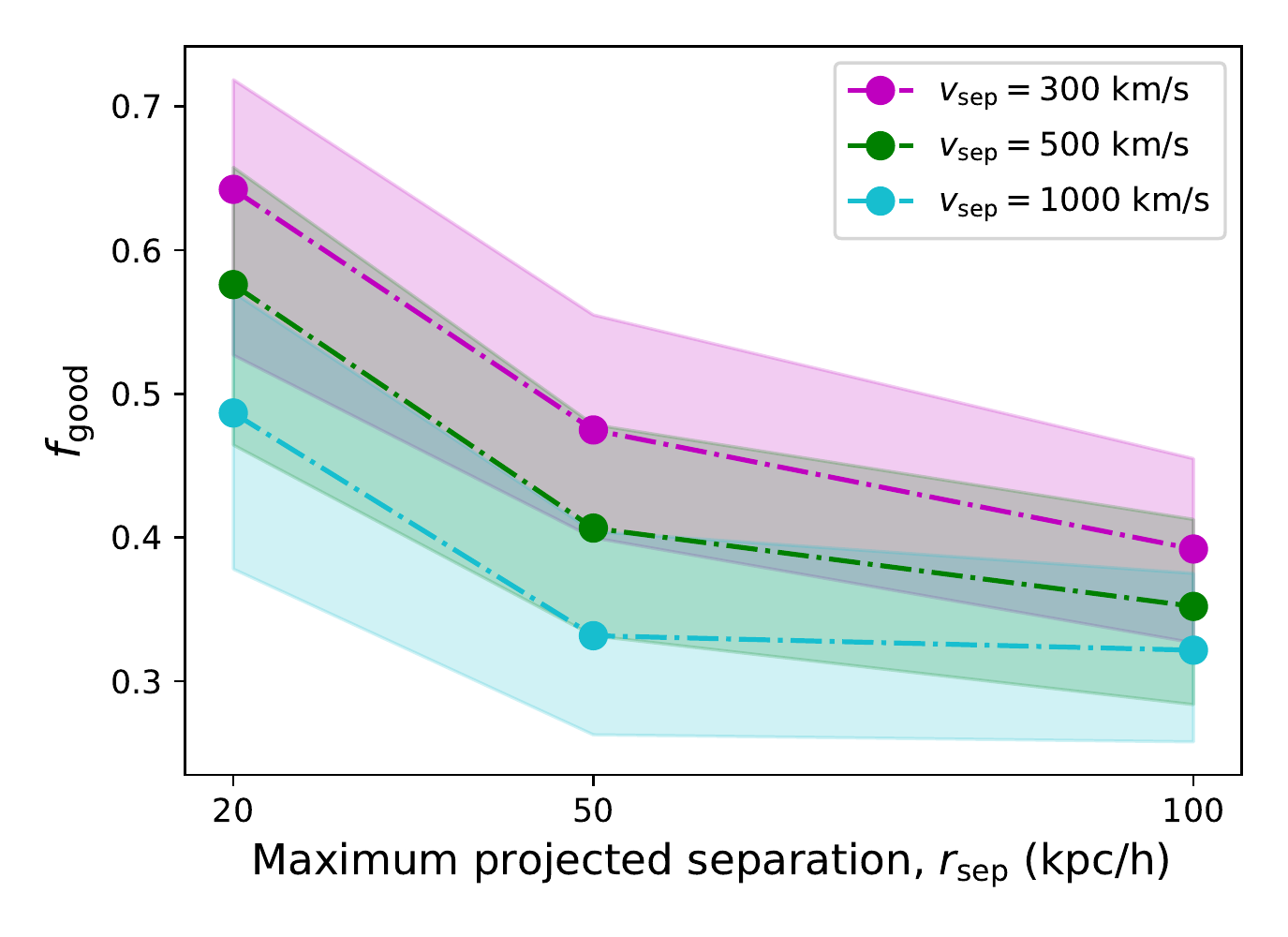}
   \caption{Same as \Fig{fig:stats2d} but for the fraction of 2D pairs that are ``good'' in 3D, as a function of maximum separation chosen, $r_\mathrm{sep}$. The dots show the median values of the distribution for the 324 clusters in \textsc{The Three Hundred} data set, with the shaded regions indicating the 16-84 percentiles. The colours indicate the velocity threshold $v_\mathrm{sep}$ used, in magenta 300 km/s, in green 500 km/s and in cyan 1000 km/s}
\label{fig:fgood}
\end{figure}

In \Fig{fig:fgood} we show this fraction for all the different values of $r_\mathrm{sep}$ and $v_\mathrm{sep}$, computed using $r_\mathrm{3D}/r_\mathrm{sep} \leq 1$ to identify good pairs. We computed this fraction separately for each cluster, so that the dots show the median value of the distribution for the 324 clusters, while the shaded regions are the 16-84 percentiles. As in \Fig{fig:stats2d}, the results are shown as a function of $r_\mathrm{sep}$ in the $x$-axis, while the different colours of the lines indicate the maximum velocity separation allowed.

We can see in the plot that the results are dependent both on the velocity and on the distance separation, with the good fraction decreasing notably for increasing values of $v_\mathrm{sep}$ and $r_\mathrm{sep}$. The difference is more significant for $r_\mathrm{sep}=20$ kpc/h, where the percentage of good pairs goes from 49 to 64 per cent for decreasing values of $v_\mathrm{sep}$. When increasing the 2D separation threshold to $r_\mathrm{sep}=50$ kpc/h, the fraction is notably reduced, ranging between 33 per cent for the highest velocity separation and 47 for the lowest one. Moving to $r_\mathrm{sep}=100$ kpc/h, the variation of the values is smaller, and they also become less dependent on the velocity threshold. For 1000 km/s, the value remains almost constant at 33 per cent, while for $v_\mathrm{sep}=500$ km/s and 300 km/s the fraction decreases to 35 and 39 per cent respectively. We also note that the scatter increases for lower values of $r_\mathrm{sep}$ due to the decreasing statistics.  

Looking at the values of the fraction in general, we have to highlight that they can be significantly low. For the most restrictive criterion, with $r_\mathrm{sep}=20$ kpc/h and $v_\mathrm{sep}=300$ km/s, the fraction of good pairs is 64 per cent, but increasing the separation threshold reduces the fraction below 50 per cent, meaning that still more than half of the close pairs observed in projections have a 3D separation in real space that is above the limit set. We will devote the next section to the study of ways to improve this result, this is, to improve the likelihood that an observed pair will be classified as good, using observable properties of the pairs.


\section{Separating the `good' from the `bad'} \label{sec:results2}
In the previous section we found that, following the methodology described in \Sec{sec:method}, the fraction of close pairs that are ``good'', according to our definition, is in general below 40 per cent. We like to remind here that our definition is only based on the 3D separation between the galaxies in the pair being below the 2D separation threshold set. 
In this section we will study some observable properties of the given pairs. By separating them into ``good'' and ``bad'' pairs, we intend to find significant differences that can be used to distinguish them. Our aim is to provide observers with an additional test to pick the ``good'' pairs from their original selection, thus improving the quality of the identified close pairs of galaxies in the sky.

\begin{figure}
  \hspace*{-0.1cm}\includegraphics[width=7.2cm]{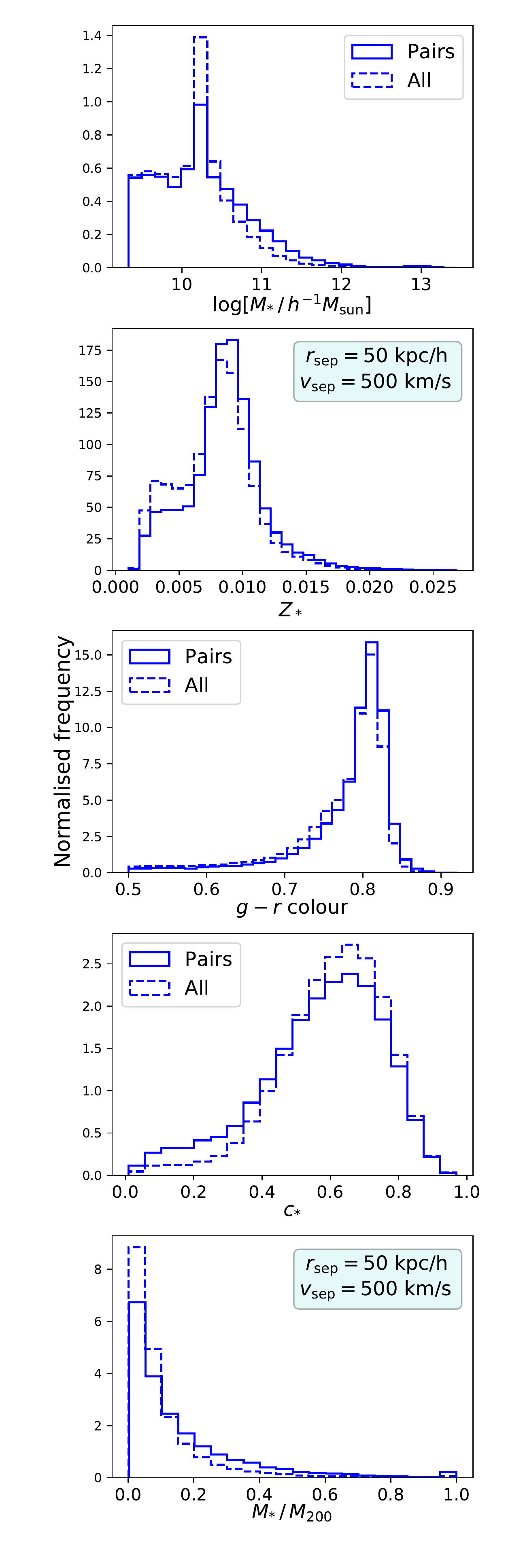}
  \caption{Comparison of the properties of galaxies in pairs (solid lines) to all the galaxies in general (dashed lines). From top to bottom:  stellar mass, stellar metallicity, $g-r$ colour, $c_\mathrm{*}$ parameter and stellar-to-halo mass ratio.}
\label{fig:props2d}
\end{figure}

\subsection{Properties of galaxies in pairs}
First of all, we explore some overall properties of the galaxies we are working with. This way, we gain some general understanding of our data set, and check that the values are within reasonable ranges. We focus on properties that are accessible to observers as well. By comparing their distribution for the galaxies in pairs as opposed to that for the whole distribution of galaxies, we can also see if there is any particular difference between them, which could be useful for further studies.

We present in \Fig{fig:props2d} the distributions of five different properties of the galaxies in pairs (solid line), together with the distributions for all the galaxies in the simulations (dashed line) that have been used throughout this work (see \Sec{sec:method} for the initial galaxy selection criteria). The first three panels in \Fig{fig:props2d} show, from top to bottom, the stellar mass of the galaxies (given by AHF), their stellar metallicity (also given in the AHF catalogues), and the $g-r$ colour (derived from the SDSS magnitudes of the galaxies that are computed using the \textsc{stardust} code, see \Sec{sec:data}). Although there are some very slight differences between the distributions, in general we see in this plot that the galaxies in pairs show no particular distribution for these properties. As before, we only show the results for one definition of proximity, with $r_\mathrm{sep}=50$ kpc/h and $v_\mathrm{sep}=500$ km/s, because the results are very similar for all of them. 

The fourth panel in \Fig{fig:props2d} shows the value of $c_{\rm *}$ defined as the ratio of the minor to major axis of the moment of inertia tensor as calculated only for the star particles. This parameter is an indicator of the shape of the galaxy, with a value of $c_\mathrm{*}$ equal to 1 indicating perfect sphericity. The bottom panel of \Fig{fig:props2d} shows the stellar-to-halo mass ratio of the galaxies in our study, which can give relevant information about the dark matter haloes of galaxies in pairs, and is computed as $M_{\rm *}/M_{200}$ using the (sub-)halo masses $M_{\rm *}$ and $M_{200}$ as given by \ahf. We can see in these plots that, although galaxies in pairs seem more likely to be less spherical and with more relevance of the stellar component than the general population, the differences between the two distributions remain minor.


\begin{figure*}
   \hspace*{-0.1cm}\includegraphics[width=13.5cm]{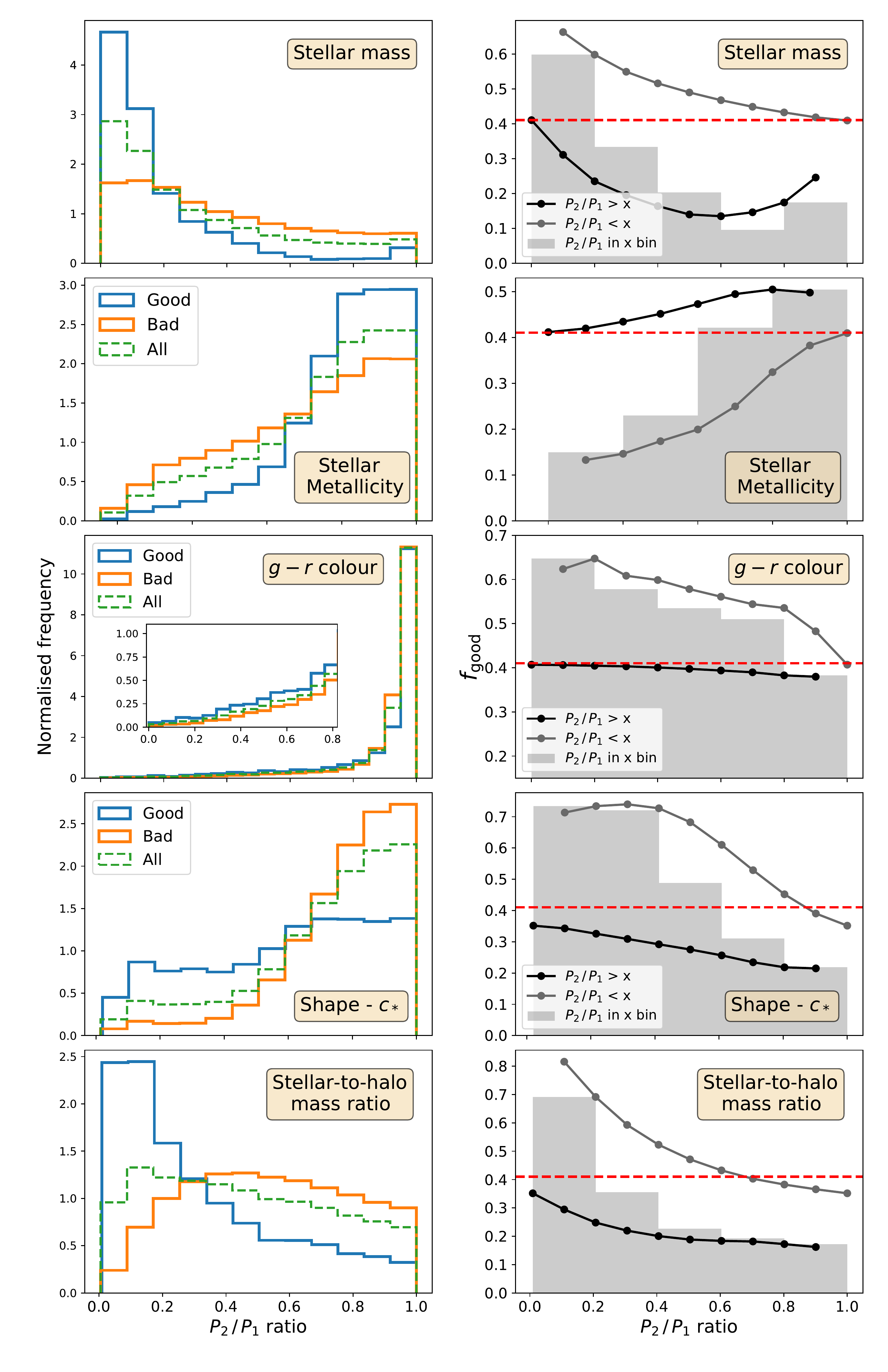}
   \caption{\textbf{Left column,} distribution of the ratio of the given property (from top to bottom: stellar mass, metallicity, $g-r$ colour, $c_*$ and stellar-to-halo mass ratio) between the two galaxies in the pair, computed for all the pairs found and separating into ``good'' and ``bad'' pairs, depending on their $r_\mathrm{3D}/r_\mathrm{sep}$ quotient. For the colour, the inset shows a zoom-in to better show the differences between the distributions. \textbf{Right column,} for the same five properties, fraction of good pairs within each bin (gray bars). The lines indicate the fraction of good pairs for the ratios being above (black) and below (gray) the given $x$-value, while the dashed horizontal line is the total good pair fraction.}
\label{fig:sep_props}
\end{figure*}

\subsection{Properties of galaxy pairs}
In \Fig{fig:props2d} we showed the properties of all the galaxies in pairs together with those for all the galaxies in our data set. Now, we are going to work instead with the properties of the galaxy pairs themselves. This is, we are going to compute, for each property, the ratio between the values of this property for both galaxies in the pair: $P_2/P_1$, where $P_2$ and $P_1$ are simply the specific properties of the two galaxies in the pair, chosen so that this ratio is always $\leq 1$. For instance, by comparing the stellar masses of the galaxies that constitute a pair, we compute their stellar mass ratio, $M_\mathrm{*2}/M_\mathrm{*1}$, which gives us information about how similar they are in mass. We compute this ratio for each pair for the five properties presented in the previous subsection.

The distributions of this ratio for all the close pairs found are shown in the left column of \Fig{fig:sep_props}. The properties are the same as in \Fig{fig:props2d}, from top to bottom: stellar mass, metallicity, colour, shape and stellar-to-halo mass ratio; although in this case we are showing the ratio between the property for the two galaxies that constitute a pair. 
Note again that this ratio is computed so that the value is always $\leq 1$. The values are computed for all the close pairs found for one definition ($r_\mathrm{sep}=50$ kpc/$h$ and $v_\mathrm{sep}=500$ km/s as before), for the 324 clusters and 100 projections, stacking them all together as previously.

In order to analyse how the probability of a pair being ``good'' is affected by either of these properties, in the left column of \Fig{fig:sep_props} we have separated the galaxies according to our definition of good presented in \Sec{sec:method}, with the results shown in \Sec{sec:results1}. The blue lines are the normalised distributions of the ratios for the good pairs, while the orange ones are for the bad pairs. The green dashed line is for all the pairs together. Each row in this figure represents the results for each of the properties in \Fig{fig:props2d}.

Starting with the first row in \Fig{fig:sep_props}, in the left panel we see that there is a clear difference between the distributions for good and bad pairs, with the former showing a more prominent peak at low values of the mass ratio. This means that good pairs tend to have a lower value than bad ones, i.e., the masses of the galaxies in good pairs are more different between each other. This is quantified in the plot in the right, which shows the fraction of good pairs for each mass ratio bin, this is, the likelihood of a pair being good given a certain value of its stellar mass ratio. The gray bars show the likelihood for the ratio being within each bin, while the black line is for a mass ratio above the given $x$-value, and the gray line for a ratio below it. The horizontal dashed line shows the overall fraction of good pairs, as shown in \Fig{fig:fgood}, so that whenever the bars or lines are above this line, this means the probability of finding a good pair is increased in that range. In the case of the stellar mass ratio, we see that for the first bin, $M_\mathrm{*2}/M_\mathrm{*1} \leq 0.2$, the likelihood increases by 46 per cent (from 41 to 60 per cent), while for ratios above this the probability decreases to 10 per cent for $0.6 < M_\mathrm{*2}/M_\mathrm{*1} \leq 0.8$.

For the second row in \Fig{fig:sep_props}, for the metallicity ratio $Z_\mathrm{*2}/Z_\mathrm{*1}$, we see the opposite difference than for the mass. Pairs that we defined as good are more likely to have a stellar metallicity ratio that is close to 1, this is, these galaxies tend to have more similar metallicities. Looking at the plot to the right to quantify this we see that, for $Z_\mathrm{*2}/Z_\mathrm{*1} > 0.8$ the likelihood is increased by $\sim 23$ per cent (from 41 to 50 per cent).

For the colour, we refer to the third row in \Fig{fig:sep_props}. The inset in the left plot shows a zoom-in for the interesting region, for ratios below 0.8. We see that the distribution here seems to be dominated by good pairs, meaning that pairs with a colour ratio below this value are more likely to be good than bad. This is confirmed in the plot beside, where we see that a colour ratio below 0.8 increases the good fraction from 41 to 54 per cent (an increase of around 30 per cent, see gray line). This likelihood increases even more when going to lower colour ratios, reaching more than 60 per cent for a ratio below 0.2, although we see in the upper panel that very few pairs are found within this range.

Regarding the two final properties, the fourth row in \Fig{fig:sep_props} shows the same distributions for $c_\mathrm{*}$. Here we can see that, while bad pairs dominate at ratios around 1, for lower values the fraction of good pairs is increased, reaching more than 70 per cent for $c_\mathrm{*2}/c_\mathrm{*1}$ < 0.4 (see right panel). This means that good pairs tend to have galaxies with different shapes, i.e. one spherical and the other one more disky, presumably due to the interaction between them. In the final row of \Fig{fig:sep_props} we study the situation for the ratio of the stellar-to-halo mass ratio between the two galaxies in the pair. We see an even clearer situation here, where good pairs prominently dominate the region of lower values in the $x$-axis, with the good pair fraction growing to almost 70 per cent for the first $x$-bin (ratios below 0.2). This can also be interpreted as an effect of the interactions between the galaxies in the pair, where one of the DM haloes is stripped by the other one, creating this difference in the stellar-to-halo mass ratio of the two involved galaxies.

We summarise our main results, as discussed in the text and extracted from \Fig{fig:sep_props}, in \Tab{table:probs}. Since  \Fig{fig:sep_props} shows the results only for $r_\mathrm{sep}=50$ kpc/h and $v_\mathrm{sep}=500$ km/s, we include in this table the results for all the different definitions of proximity used throughout the paper, for the three different values for the parameters $r_\mathrm{sep}$ and $v_\mathrm{sep}$. The first column in this table simply shows the fraction of good pairs, although, unlike in \Fig{fig:fgood}, it is computed by stacking all the pairs for all the clusters together, rather than computing $f_\mathrm{good}$ for each cluster and then getting the median value for the 324 clusters. The following columns indicate by which amount is this fraction increased when only the pairs within certain ranges of the studied properties are considered. As an example, for the first row, the fraction of good pairs increases by 21.4 per cent when limiting the pairs to those with $M_\mathrm{*2}/M_\mathrm{*1} \leq 0.2$. This means that it goes from $f_\mathrm{good}=0.642$ (see first column) to $f_\mathrm{good} \times (1+0.214) = 0.780$. 
In this table we see that the overall situation is similar for the different values of $r_\mathrm{sep}$ and $v_\mathrm{sep}$, the main difference being for $r_\mathrm{sep}=20$ kpc/h, where the colour and $c_*$ ratios (fourth and fifth columns) do not separate good and bad pairs for the given range, especially for the two lower $v_\mathrm{sep}$ value, where the negative sign means the probability decreases. We like to note here that, although selecting a lower threshold in the colour ratio would mean a greater increase in the likelihood (see \Fig{fig:sep_props}), it can also be seen in this figure that the number of pairs is very low for these bins, and so we prefer to use 0.8 as the threshold for the table. For the shape parameter, $c_*$, we can see in the table that the results are very dependent on the maximum 2D separation allowed and, while in the given range the likelihood slightly decreases for $r_\mathrm{sep}=20$ kpc/h, it can be increased by more than 100 per cent for $r_\mathrm{sep}=100$ kpc/h.

We conclude from this subsection that, using these five observable properties and computing the ratio for an observed pair, its value can be used as an indicator of the probability of that pair being good according to our definition. To further increase the likelihoods, the criteria for the different properties can be combined. For instance, for $r_\mathrm{sep}=50$ kpc/h and $v_\mathrm{sep}=500$ km/s, if a pair has both mass ratio below 0.2 and metallicity ratio above 0.8, the likelihood of it being good increases from the initial 41 to more than 75 per cent (higher than for each of them individually). For simplicity, we do not include these numbers in the table, but we note that every combination of them results in an ever increased percentage of good pairs.

\begin{table*}
\centering
\caption{For each $r_\mathrm{sep}$ and $v_\mathrm{sep}$ combination, fraction of good pairs (``All pairs'' column) and amount by which this fraction is increased when considering only the pairs within the given range for each property (see second column in \Fig{fig:sep_props} for the values for 50 kpc/h and 500 km/s). Note that the values in the first column, $f_\mathrm{good}$, are not exactly the same as in \Fig{fig:fgood}, since for this table they are computed as the fraction of good pairs for all the projections for all the clusters stacked together; while in \Fig{fig:fgood} the value for each cluster was computed and then the median values (and 16-84 percentiles) for the 324 clusters are shown.}
\begin{tabular}{r|c|c|c|c|c|c|c}
\hline
             &                 & $\bm{f_\mathrm{good}}$ & \multicolumn{5}{c}{\textbf{Increase in} $\bm{f_\mathrm{good}}$} \\
\multicolumn{1}{r}{$\bm{r_\mathrm{sep}}$\textbf{(kpc/h)}} & $\bm{v_\mathrm{sep}}$\textbf{(km/s)}  & {All pairs} & {$M_\mathrm{*}$ ratio} ${\leq 0.2}$ & {$Z_\mathrm{*}$ ratio} ${\geq 0.8}$ & {$g-r$ ratio} ${\leq 0.8}$ & {$c_*$ ratio} ${\leq 0.4}$ & {SMHM ratio} ${\leq 0.2}$ \\ \hline \hline
\textbf{20}  & \textbf{300}    & 0.642    & 0.214     & 0.076     & -0.086     & -0.094    & 0.279         \\ 
             & \textbf{500}    & 0.576    & 0.281     & 0.105     & -0.035     & -0.019    & 0.374          \\ 
             & \textbf{1000}   & 0.484    & 0.400     & 0.153     & 0.047      & 0.121     & 0.548          \\ \hline 
\textbf{50}  & \textbf{300}    & 0.481    & 0.366     & 0.188     & 0.220      & 0.332     & 0.622           \\ 
             & \textbf{500}    & 0.410    & 0.458     & 0.230     & 0.305      & 0.520     & 0.791          \\ 
             & \textbf{1000}   & 0.336    & 0.572     & 0.285     & 0.404      & 0.787     & 1.023          \\ \hline 
\textbf{100} & \textbf{300}    & 0.392    & 0.437     & 0.244     & 0.366      & 0.883     & 0.817          \\ 
             & \textbf{500}    & 0.352    & 0.492     & 0.276     & 0.379      & 1.068     & 0.967           \\ 
             & \textbf{1000}   & 0.318    & 0.538     & 0.317     & 0.346      & 1.248     & 1.126           \\ \hline
\end{tabular}
\label{table:probs}
\end{table*}

\subsection{Dependence of the separation on distance to cluster centre}
\begin{figure}
  \hspace*{-0.1cm}\includegraphics[width=8cm]{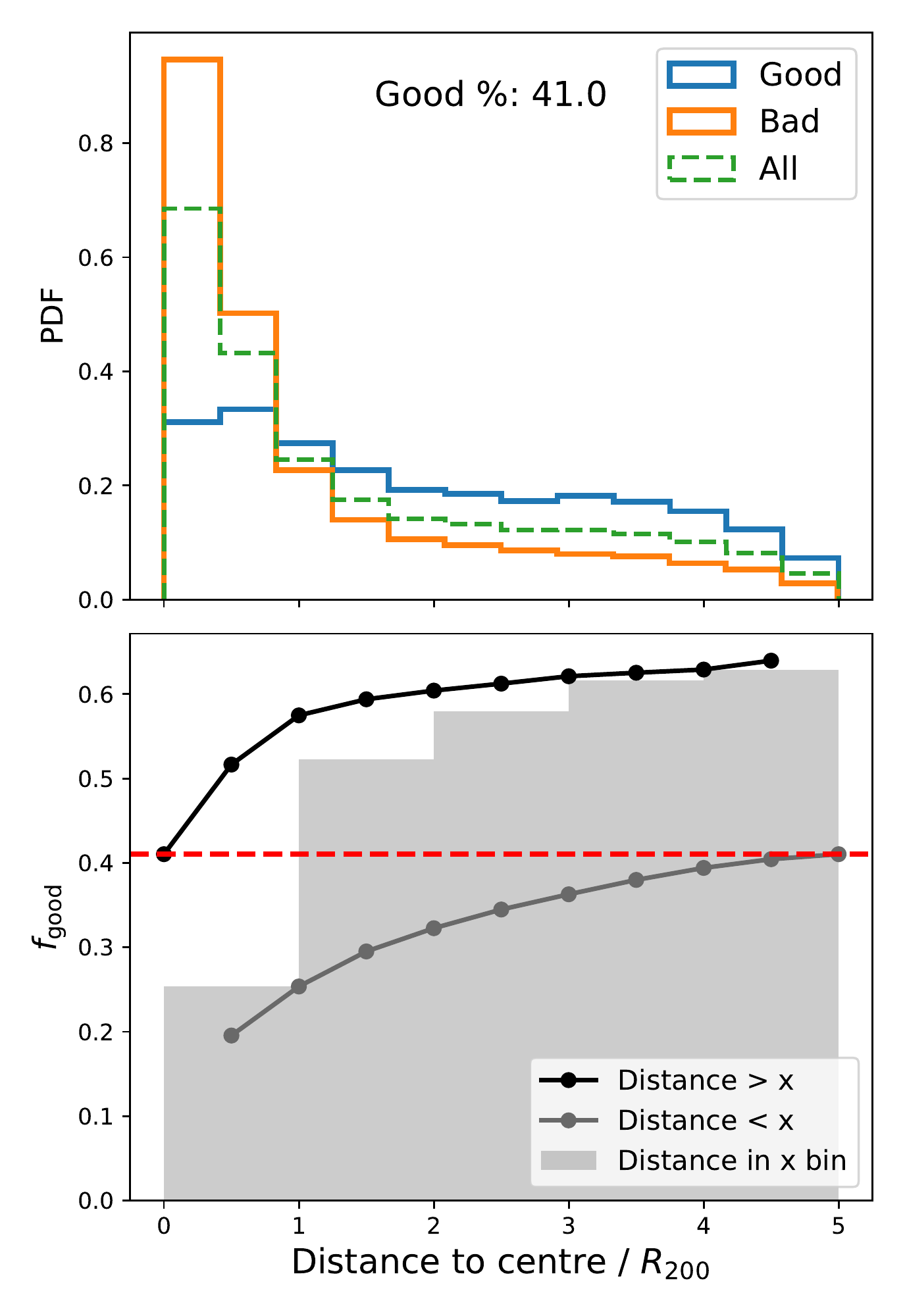}
  \caption{Same as \Fig{fig:sep_props} but for the 2D distance from the pair to the cluster centre, in units of $R_{200}$. \textbf{Top,} distribution for all the pairs separated into ``good'' and ``bad'' pairs, depending on the quotient $r_\mathrm{sep}$/3D distance ratio. \textbf{Bottom,} fraction of good pairs within each bin (gray bars), and above (black line) and below (gray line) the given $x$-value. The dashed horizontal line is the total good pair fraction.}
\label{fig:sep_dist2cen}
\end{figure}

Apart from the properties previously discussed, there is another property of the pairs that is easy to check observationally, and that can give interesting information about the pair. This is the projected distance to the cluster centre. We measure this distance for all the pairs, and represent it in units of the cluster radius $R_{200}$ in \Fig{fig:sep_dist2cen}. In this figure we basically repeat the plot in \Fig{fig:sep_props} but only for this distance to centre.

In the upper panel of \Fig{fig:sep_dist2cen} we see that there is a very clear difference between good and bad pairs. While the good ones are almost uniformly distributed along the $5R_{200}$ zone, the bad ones are preferentially located inside the cluster. This is something that could be expected, the higher density of galaxies towards the centre makes it harder to identify real pairs. 
However, it is still an interesting result, that already warns that galaxy pairs close to the cluster centre are much more likely to be farther away in physical distance. 
In fact, looking at the lower panel in this figure, we can see that the likelihood of finding a good pair inside $R_\mathrm{200}$ decreases to 25 per cent, while it grows to 57 per cent for distances above this radius. 

\begin{figure*}
  \hspace*{-0.1cm}\includegraphics[width=13cm]{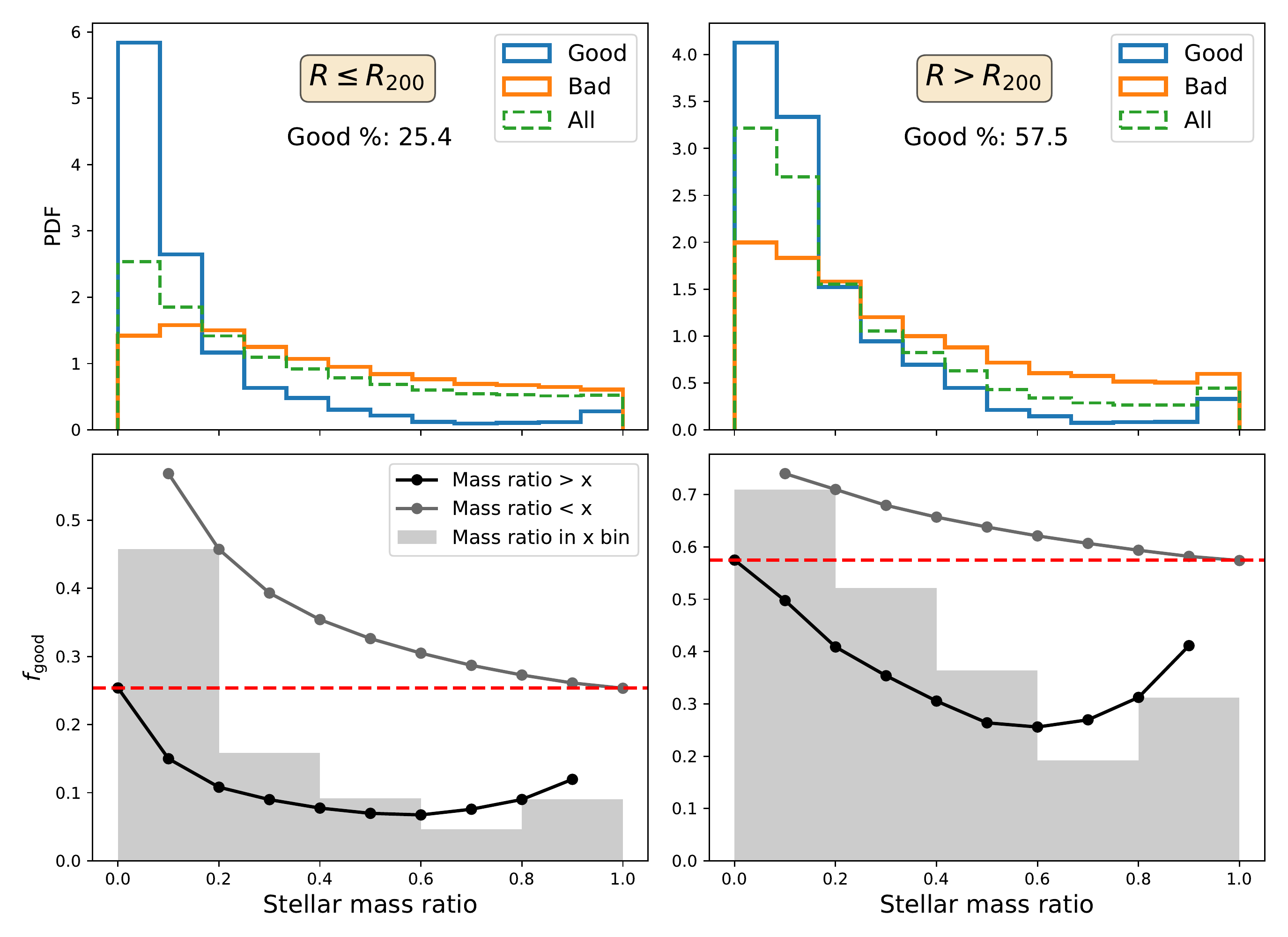}
  \caption{Same as first row in \Fig{fig:sep_props} but separating into the region inside the cluster (left) and outside the cluster (right).}
\label{fig:sep_massinout}
\end{figure*}

Now, given this difference in the number of good pairs depending on the region we are in, we ask ourselves whether the results obtained in the previous subsection are just an effect of this result regarding the distance to the cluster centre. This is, we question if the differences in mass ratio, colour and metallicity between good and bad pairs are just a consequence of these pairs being located in different regions of the cluster environment. To answer this question, in \Fig{fig:sep_massinout} we present the same plot as the first row in \Fig{fig:sep_props} but separating for the region inside the cluster, $R \leq R_{200}$ (left), and outside the cluster, $R > R_{200}$ (right). 

We see that, although the results weaken, the differences still hold outside the cluster, with good pairs having in general lower values of the stellar mass ratio. In the bottom left panel we see that, in the $M_\mathrm{*2}/M_\mathrm{*1} \leq 0.2$ bin, the probability increases from 25 to 46 per cent, while in the right one the increase is from 57 to 71 per cent in this same bin.
Although not shown here, we repeated these plots for the other four different properties analysed throughout the paper, finding that the trends seen always hold in both regions inside at outside the cluster. For the metallicity and the colour ratios, a similar situation to the mass is found, with the results weakening a bit outside the cluster, while for the shape and the stellar-to-halo mass ratio, the results remain the same regardless of the region.

We note that, when talking about the cluster centre, we refer to the maximum density peak of the main halo, identified by our halo finder, rather than taking a more observational approach like using the position of the brightest cluster galaxy (BCG). Nevertheless, several works have already shown that the offset between these two positions is small, especially in relaxed clusters, so the same results are expected if using the BCG position or the centroid of X-ray emission (see \citealp{Cui2016} for an in-depth study on how the centre position is affected by the method chosen to identify it; or \citealp{DeLuca2021}, where the offset is computed for \textsc{The Three Hundred} clusters).

\subsection{Dependence of the separation on cluster properties}
In the previous subsection we tried to analyse the probability of correctly identifying a galaxy pair in a 2D observation as a function of the properties of the pairs themselves. Now, we also want to study if there is a correlation between this probability of success and the properties of the whole cluster. 

\begin{figure}
   \hspace*{-0.1cm}\includegraphics[width=8.5cm]{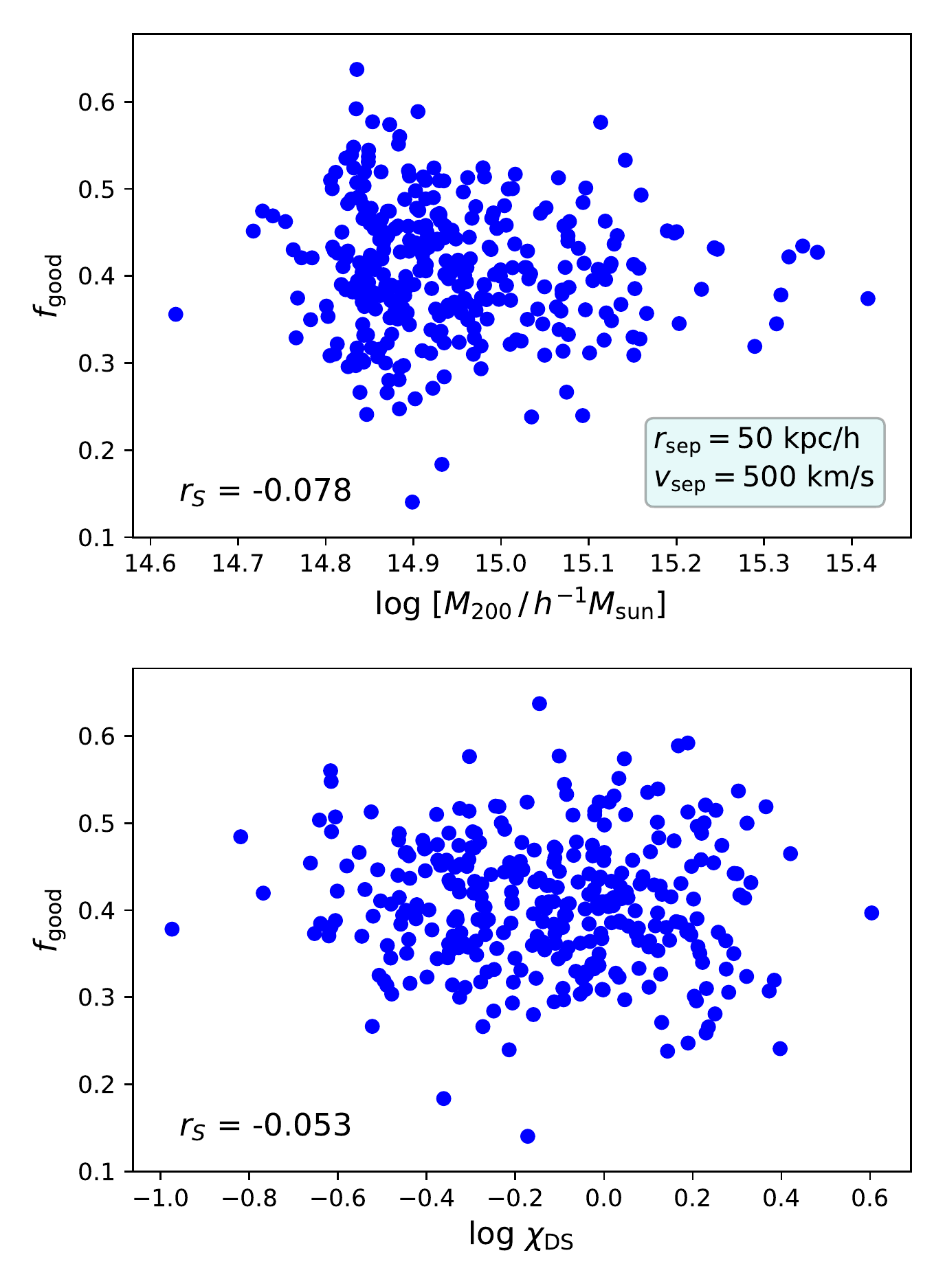}
   \caption{\textbf{Top}, fraction of good pairs of galaxies, computed for $r_\mathrm{sep}=50$ kpc/h and $v_\mathrm{sep}=500$ km/s, with the selection criterion $r_\mathrm{good}=0.8$ for good pairs, as a function of mass of the cluster, $M_{200}$, in logarithmic scale.
  \textbf{Bottom}, same good fraction as top panel, but as a function of the relaxation parameter $\chi_\mathrm{DS}$. In both panels, the Spearman correlation coefficient, $r_S$, is indicated in the bottom left corner.}
\label{fig:fgood_corr}
\end{figure}

In order to do this, we tried to find a correlation between the fraction of pairs that were correctly identified for each cluster, $f_\mathrm{good}$, as shown in \Fig{fig:fgood}, and any cluster property. We show this in \Fig{fig:fgood_corr}, for the total halo mass of the cluster, $M_{200}$, and for the cluster's dynamical state, quantified by the relaxation parameter $\chi_\mathrm{DS}$, as explained in \Sec{sec:stats_corr}. In both cases we could not find any significant correlation, meaning that the probability of an observed pair being good does not depend on properties of the cluster as a whole but rather of those of the pair itself. 
In \Sec{sec:stats_corr} we found a significant correlation between the fraction of galaxies in a pair or group and the dynamical state of the cluster, with disturbed clusters having more pairs and groups. Now we see that the fraction of good pairs is not dependent on the dynamical state of the cluster, so for unrelaxed clusters we find more good pairs but also more bad pairs, meaning that projection effects affect relaxed and unrelaxed clusters equally.

\section{Conclusions} \label{sec:conclusions}

In this work we studied close pairs of galaxies, that are used in observations as an indicator of galaxy mergers and interactions. We worked with a set of cosmological simulations of galaxy clusters and their surroundings, allowing us to additionally investigate any correlation with environment. By taking advantage of the power of simulations, we evaluated the performance of the observational techniques used to identify these pairs, and proposed ways, using observable properties of the galaxies, to improve this performance.

The simulations used in this work are provided by \textsc{The Three Hundred} project, and consist of a set of 324 numerically modelled spherical  regions centred on the most massive clusters found in the  DM-only MDPL2 MultiDark Simulation. 
For each of these simulations, we limit our study to the region within $5R_{200}$ of the cluster halo centre, and select all the galaxies with $M_\mathrm{*} > 10^{9.5} \hMsun$. 
To mimic observational studies, we randomly rotate the 3D coordinates and then project them into a 2D plane, so that we can get numerous projections of the same cluster. 
We applied the same techniques used by observers to find close pairs of galaxies. This way, we identified close pairs as those that lie within a separation in the sky smaller than $r_\mathrm{sep}$ and a separation in velocity along the line of sight below $v_\mathrm{sep}$ (where we included the Hubble flow). The values used for $r_\mathrm{sep}$ (20, 50 and 100 kpc/h) and for $v_\mathrm{sep}$ (300, 500 and 1000 km/s) were based on previous studies in the literature, and combined for a broad research of close pairs and the dependence on these parameters. We also allowed galaxies to have more than one close companion, and hence being in groups.

After grouping the galaxies for the different clusters and projections, in \Fig{fig:stats2d} we analysed the statistics of the results. We showed first the fraction of galaxies having at least one close companion, that depends a lot on the $r_\mathrm{sep}$ threshold used. We saw that our results are overall in line with previous observational results. We also showed in \Fig{fig:stats2d} that the great majority of the connected galaxies are in pairs rather than groups, with the latter becoming more relevant for $r_\mathrm{sep}=100$ kpc/h. For this reason, the following parts of our work were only focused on galaxy pairs, that also allow for a simpler analysis.

The next step in this study was to use the physical coordinates available in the simulations to obtain the 3D separations between the galaxies in pairs. We computed the ratio $r_\mathrm{3D}/r_\mathrm{sep}$ for all the pairs found, where $r_\mathrm{sep}$ is the fixed threshold set for the projected separation in the sky for each definition of proximity. This way, we could divide the pairs into ``good'' pairs: those with $r_\mathrm{3D}/r_\mathrm{sep} \leq 1$, i.e., with a 3D separation within the allowed $r_\mathrm{sep}$ range; and ``bad'' pairs, which do not satisfy this condition. 
Using this threshold, in \Fig{fig:fgood} we depicted the fraction of good pairs for all the different definitions. This fraction is shown to be between 50 and 65 per cent for the smallest $r_\mathrm{sep}$ value used, while it drops significantly as this parameter increases. For maximum separations of 50 and 100 kpc/h, the fraction of good pairs ranges between 30 and less than 50 per cent depending on the velocity separation limit.

For a better understanding of this separation between good and bad pairs, we devoted \Sec{sec:results2} to studying observable properties of these pairs and the galaxies within them, trying to find differences that can be used to distinguish them. We worked with the stellar mass, stellar metallicity and colour. We also studied the shape, quantified by $c_*$, and stellar-to-halo mass ratio. In \Fig{fig:sep_props} we analysed the ratio of each property between the two galaxies in every pair, distinguishing the values for good and bad pairs. Regarding the stellar mass ratio, we found that good pairs tend to have a lower value, i.e. the galaxies are more different in mass in these pairs. The situation is reversed for metallicity, where a ratio close to 1 indicates that the pair is more likely to be a good one. Concerning the colours, \Fig{fig:sep_props} shows that for ratios below $\sim 0.8$ the pairs are also more likely to be good. These results are quantified, and summarised for each $r_\mathrm{sep} - v_\mathrm{sep}$ combination, in \Tab{table:probs}. We saw in this table that, generally speaking, the fraction of good pairs is increased by 30-50 per cent if the given galaxy pair has $M_\mathrm{*2}/M_\mathrm{*1} \leq 0.2$, $Z_\mathrm{*2}/Z_\mathrm{*1} \geq 0.8$ or a $g-r$ colour ratio below 0.8, although these results weaken if the 2D separation threshold to find the pairs is set at 20 kpc/h. 
For the shape and stellar-to-halo mass ratio, we found in both cases that good pairs tend to have lower values, again meaning that, regarding these properties, the situation for the two galaxies involved in the pair is significantly different. This can be interpreted as a consequence of the interaction between the galaxies. \Tab{table:probs} shows that, specially for 50 and 100 kpc/h separations, selecting lower values within these properties can increase the fraction of good pairs by 50-100 per cent.

Apart from these properties, we also studied the projected distance from the pair to the cluster centre. We found that separating the good from the bad pairs also gives a clear difference in the distributions, with the bad pairs more clearly found near the cluster centre, within $R_{200}$ (see \Fig{fig:sep_dist2cen}). However, we found that the results for the previous properties, although weakened in some cases (for the stellar mass, metallicity and colour ratios), still hold if we separate for the regions inside and outside $R_{200}$ (see \Fig{fig:sep_massinout} for the mass), so they are not just a consequence of bad pairs being more predominant in the inner region. 
In this regard, \citet{Ellison2010} investigated galaxy pairs as a function of the environment, finding that, although galaxy interactions seem to be ubiquitous in the Universe, their observational manifestations show a dependence on the environment.
Here we like to note again that our study has been done for galaxy pairs in cluster environments, reaching up to $5R_{200}$ of the main cluster centre (additional massive objects can also be found here), so that our results are only valid for these regions and they may not hold for randomly selected pairs. Future works repeating this study on a different environment could provide with more information regarding how the results are affected by this.

Previous works have already used simulations to similarly analyse observational results. For instance, \citet{McConnachie2008} used mock galaxy catalogues based on the Millennium simulation to investigate the spatial properties of groups of galaxies, finding that only $\sim 30$ per cent of the groups were truly compact in three dimensions. In a later work by the same authors, \citet{McConnachie2009} showed that selecting groups by surface brightness reduces this effect significantly, increasing the fraction of physically compact groups.
Regarding the properties of the involved galaxies, several works have studied how they are affected by close companions and interactions. \citet{Scudder2012} studied galaxy pairs in the Sloan Digital Sky Survey, tracing the changes induced in star formation rates and metallicities. \citet{Alonso2006} also studied star formation and colour of the galaxies in observed pairs, and how they are affected by interactions. 
In this work we take a different approach, based on observational properties of the pairs rather than the galaxies, i.e. using the ratio $P_2/P_1$, where $P_2$ and $P_1$ are the values of the given property for the two galaxies. We propose the use of these ratios to increase the likelihood of a pair being close in physical distance too, and thus reduce these effects that deteriorate close-pair catalogues.

In conclusion, our results show that, computing the two galaxies' ratio for the projected pairs, the stellar mass, metallicity, $g-r$ colour, shape and stellar-to-halo mass ratio can be used as indicators of the probability of the pairs being good according to our definition. This is simply based on the 2D distance threshold used, $r_\mathrm{sep}$, and the 3D separation between the galaxies, $r_\mathrm{3D}$.
We note again that this definition of ``good'' is only a measure of 3D proximity, rather than physically testing if the galaxies are bound. Previous works like \citet{Haggar2021} or \citet{Choque-Challapa2019} associate galaxies with a host galaxy based on their relative velocity and the host gravitational potential. In future works following this study we aim at using a more theoretical approach to distinguish between these physical pairs and galaxies that are just passing by. Besides, although in this paper we mainly focused on pairs of galaxies, a more in-depth study could be done regarding groups of galaxies and observational techniques to find them. A similar work has also been performed recently by \citet{Taverna2022}, analysing the completeness and reliability of compact group catalogues using semi-analytical models of galaxy formation.


\section*{Acknowledgements}
\addcontentsline{toc}{section}{Acknowledgements}

This work has been made possible by \textsc{The Three Hundred} (\url{https://the300-project.org}) collaboration. The simulations used in this paper have been performed in the MareNostrum Supercomputer at the Barcelona Supercomputing Center, thanks to CPU time granted by the Red Espa\~{n}ola de Supercomputaci\'on. As part of \textsc{The Three Hundred} project, this work has received financial support from the European Union’s Horizon 2020 Research and Innovation programme under the Marie Sklodowskaw-Curie grant agreement number 734374, the LACEGAL project. AC, AK, and GY are supported by the MICIU/FEDER through grant number PGC2018-094975-C21 as well as by the Ministerio de Ciencia e Innovaci\'{o}n under research grant PID2021-122603NB-C21. AK further thanks Urge Overkill for sister havanna. WC is supported by the STFC AGP Grant ST/V000594/1 and by the Atracción de Talento Contract no. 2020-T1/TIC-19882 granted by the Comunidad de Madrid in Spain. He further acknowledges the science research grants from the China Manned Space Project with NO. CMS-CSST-2021-A01 and CMS-CSST-2021-B01. RH acknowledges support from STFC through a studentship.

\section*{Data Availability}
The results shown in this work use data from \textsc{The Three Hundred} galaxy clusters sample. These data are available on request following the guidelines of \textsc{The Three Hundred} collaboration, at \url{https://www.the300-project.org}. The data specifically shown in this paper will be shared upon request to the authors.



\clearpage
\bibliographystyle{mnras}
\bibliography{archive}

\appendix


\bsp	
\label{lastpage}
\end{document}